\PassOptionsToPackage{format=plain,labelformat=default, labelsep=endash, labelfont=bf, margin=0pt, tableposition=top, font={footnotesize}}{caption}
\PassOptionsToPackage{usenames}{color}
\documentclass[twocolumn]{aastex} 

\usepackage{graphicx}  
\usepackage{amssymb}
\usepackage{array,multirow}
\usepackage[T1]{fontenc}
\usepackage{ae,aecompl}
\usepackage{amsmath}	
\usepackage{subfigure}
\usepackage{listings}
\usepackage{xspace}
\usepackage{url}
\usepackage{times}
\usepackage[english]{babel}
\usepackage{mathrsfs}
\usepackage{threeparttable}
\usepackage{booktabs}
\newcommand{\ie}{i.\,e.} 
\newcommand{\eg}{e.\,g.} 

\usepackage{twoopt}
\usepackage[varg]{txfonts}
\usepackage[normalem]{ulem}

\newcommand{\fref}[1]{Figure~\ref{#1}}
\newcommand{\tref}[1]{Table~\ref{#1}}
\newcommand{\sref}[1]{Section~\ref{#1}}

\newcommand{\pipe}{\ensuremath{\rm K2P^2}\xspace}

\newcommand{\numax}{\ensuremath{\nu_{\rm max}}\xspace}
\newcommand{\dnu}{\ensuremath{\Delta\nu}\xspace}
\newcommand{\kp}{\emph{Kepler}\xspace}
\newcommand{\Kp}{\ensuremath{\rm Kp}\xspace}
\newcommand{\teff}{\ensuremath{T_{\rm eff}}\xspace}
\newcommand{\feh}{\ensuremath{\rm [Fe/H]}\xspace}

\makeatletter
\def\maketag@@@#1{\hbox{\m@th\normalfont\normalsize#1}}
\makeatother

\bibpunct{(}{)}{;}{a}{}{,} 

\makeatletter
\newcommand*\mysize{%
  \@setfontsize\mysize{5.8}{7}%
}
\makeatother

\makeatletter
\newcommand*\tabsize{%
  \@setfontsize\tabsize{7.}{8.0}%
}
\makeatother

\begin{document}

\slugcomment{Accepted for Publications of the Astronomical Society of the Pacific (PASP)}

\title{Asteroseismic properties of solar-type stars observed with the NASA K2 mission: results from Campaigns 1-3 and prospects for future observations}
\shorttitle{Asteroseismology with K2}

\author{Mikkel~N.~Lund\altaffilmark{1,2}$^{\star}$}\email{$^{\star}$lundm@bison.ph.bham.ac.uk}
\author{William~J.~Chaplin\altaffilmark{1,2}}
\author{Luca~Casagrande\altaffilmark{3}}
\author{V{\'{\i}}ctor~Silva~Aguirre\altaffilmark{2}}
\author{Sarbani~Basu\altaffilmark{4}}
\author{Allyson Bieryla\altaffilmark{5}}
\author{J\o rgen~Christensen-Dalsgaard\altaffilmark{2}}
\author{David~W.~Latham\altaffilmark{5}}
\author{Timothy~R.~White\altaffilmark{2}}
\author{Guy~R.~Davies\altaffilmark{1,2}}
\author{Daniel~Huber\altaffilmark{6,7,2}}
\author{Lars~A.~Buchhave\altaffilmark{8}}
\and \author{Rasmus~Handberg\altaffilmark{2}}

\altaffiltext{1}{School of Physics and Astronomy, University of Birmingham, Edgbaston, Birmingham, B15 2TT, UK}
\altaffiltext{2}{Stellar Astrophysics Centre, Department of Physics and Astronomy, Aarhus University, Ny Munkegade 120, DK-8000 Aarhus C, Denmark}
\altaffiltext{3}{Research School of Astronomy and Astrophysics, Mount Stromlo Observatory, The Australian National University, ACT 2611, Australia}
\altaffiltext{4}{Department of Astronomy, Yale University, PO Box 208101, New Haven, CT 06520-8101, USA}
\altaffiltext{5}{Harvard-Smithsonian Center for Astrophysics, 60 Garden Street Cambridge, MA 02138 USA}
\altaffiltext{6}{Sydney Institute for Astronomy (SIfA), School of Physics, University of Sydney, NSW 2006, Australia}
\altaffiltext{7}{SETI Institute, 189 Bernardo Avenue, Mountain View, CA 94043, USA}
\altaffiltext{8}{Centre for Star and Planet Formation, Natural History Museum of Denmark \& Niels Bohr Institute, University of Copenhagen, \O ster Voldgade 5-7, DK-1350 Copenhagen K, Denmark}

\received{2016 May 3}
\revised{2016 August 9}
\accepted{2016 August 22}

\shortauthors{Mikkel~N.~Lund et al.}

\begin{abstract}
We present an asteroseismic analysis of 33 solar-type stars observed in short cadence during Campaigns (C) 1-3 of the NASA K2 mission. We were able to extract both average seismic parameters and individual mode frequencies for stars with dominant frequencies up to ${\sim}3300\,\rm\mu Hz$, and we find that data for some targets are good enough to allow for a measurement of the rotational splitting. 
Modelling of the extracted parameters is performed by using grid-based methods using average parameters and individual frequencies together with spectroscopic parameters. 
For the target selection in C3, stars were chosen as in C1 and C2 to cover a wide range in parameter space to better understand the performance and noise characteristics. For C3 we still detected oscillations in $73\%$ of the observed stars that we proposed.
Future K2 campaigns hold great promise for the study of nearby clusters and the chemical evolution and age-metallicity relation of nearby field stars in the solar neighbourhood. 
We expect oscillations to be detected in ${\sim}388$ short-cadence targets if the K2 mission continues until C18, which will greatly complement the ${\sim}500$ detections of solar-like oscillations made for short-cadence targets during the nominal \kp mission. For ${\sim}30-40$ of these, including several members of the Hyades open cluster, we furthermore expect that inference from interferometry should be possible. 
\end{abstract}

\keywords{Asteroseismology --- methods: data analysis --- stars: solar-type --- stars: oscillations --- stars: fundamental parameters --- stars: distances}


\section{Introduction}
The study of solar-type stars by virtue of asteroseismology has been one of the great successes of the NASA \kp mission \citep[][]{2010PASP..122..131G}.
These studies include both ensemble analysis of field stars \citep[][]{2014ApJS..210....1C}, and inferences on planet hosting stars \citep[][]{2013ApJ...767..127H,2014A&A...570A..54L,2014ApJ...782...14V,2015ApJ...799..170C}, including detailed analysis from individual mode frequencies \citep[][]{2015MNRAS.452.2127S,2016MNRAS.456.2183D}. 

The loss of a second of \kp's four reaction wheels ended the nominal mission in May of 2013. With good grace, \kp was expertly repurposed by the mission teams into the ecliptic plane K2 mission \citep[][]{2014PASP..126..398H}. With its observations along the ecliptic plane K2 offers a unique opportunity to study different regions of the Galaxy. Data from Campaign 1 (C1) have already offered asteroseismic results of solar-like and red-giant field stars \citep[][]{2015PASP..127.1038C,2015ApJ...809L...3S}. From C3 onwards the operation of the fine-guidance sensors changed, resulting in a better pointing and a significantly improved high-frequency performance \citep[][]{2016van.cleve.pasp} --- with the improved understanding of the data characteristics and noise properties the prospects for future results are promising.

Compared to the nominal \kp mission, K2 will observe many nearby bright stars, which will give us the opportunity to apply asteroseismology to study the local solar neighbourhood, placing constraints on the age-metallicity relation of nearby field stars.
The bright targets bring powerful tests of stellar structure and evolution to bear, because follow-up and complementary observations (spectroscopy, astrometry, and interferometry) may be more readily obtained and combined with asteroseismology. This will also allow a better calibration of seismic scaling relations, which are important to asteroseimic ensemble studies for galactic archaeology \citep[][]{2013MNRAS.429..423M,2014ApJ...787..110C,2016MNRAS.455..987C,2015MNRAS.449.2604D,2016ApJ...822...15S}.  
Moreover, K2 allows us to study many interesting stellar clusters, which were only sparsely represented in the nominal \kp mission, especially young clusters --- these include the Hyades (Lund et al., in press), the Pleiades (White et al., in prep.), Praesepe/M44, and M67 (Stello et al. in prep.), in addition to the old globular cluster M4 \citep[][]{2016MNRAS.461..760M}.

It is important to note that a similar view of the Galaxy will not become available with the missions coming online in the near future; the Transiting Exoplanet Survey Satellite \citep[TESS;][]{2014SPIE.9143E..20R,2015JATIS...1a4003R} will perform a near-all-sky survey, but it will largely avoid the ecliptic. PLATO \citep[][]{2013arXiv1310.0696R} will observe longer and will likely sample several diverse fields in its step-and-stare phase, but data will first become available after 2024. For both these missions the study of clusters will be challenging because of the larger pixel sizes adopted.   

In this paper we demonstrate the utility of K2 for carrying out asteroseismic studies of field stars in the solar neighbourhood. We find that the photometric quality of the data from C3 is better than expected, and we detect oscillations in $73\%$ of the observed stars that we proposed. With the extracted seismic parameters, comprising both average quantities and individual frequencies, we are able to perform grid-based and detailed seismic modelling. 

The paper is organised as follows: in \sref{sec:data} we present the reduction of K2 data, and the atmospheric parameters obtained for the targets with detected oscillations. \sref{sec:param} describes the seismic parameters; \sref{sec:model} is devoted to the modelling of these, using several different pipelines. The prospect for future K2 campaigns is the subject of \sref{sec:fut}, where we look at the question of noise characteristics, our ability to detect oscillations, and we assess the possibility of complementary interferometric observations. We end with concluding remarks in \sref{sec:dis}.


\section{Data}
\label{sec:data}

We have analysed targets observed in short cadence (SC) during K2 campaigns 1-3, which were included in the guest-observer (GO) proposals 1038, 2038, and 3038 (see \tref{tab:fut_cam}).
In total 88 targets were selected for observation. The primary selection criteria for the targets were based on a predicted detectability of solar-like oscillations in the frequency range suited for SC observations. Another concern was to choose stars that cover a wide range in parameter space to obtain a better understanding of the performance and noise characteristics. For details see \citet[][]{2015PASP..127.1038C}.

The K2-Pixel-Photometry pipeline \citep[\pipe;][]{2015ApJ...806...30L} was used to extract light curves from background-corrected pixel data\footnote{downloaded from the KASOC database; \url{www.kasoc.phys.au.dk}}. For the targets analysed here, all of which are saturated, we defined custom pixel masks. We then corrected the light curves for instrumental trends from the apparent ${\sim}6$-hour motion of the targets on the CCD, using the KASOC filter (\citet[][]{2014MNRAS.445.2698H}; see also Handberg \& Lund, in press). 
Weighted power density spectra (PDS) used for extracting the seismic parameters were created using a least-squares sine-wave fitting method, similar to the classical Lomb-Scargle periodiogram \citep[see][]{1976Ap&SS..39..447L,1982ApJ...263..835S}, but with the addition of statistical weights \citep{1992PhDT.......208K,1995A&A...301..123F}. The PDS were normalised by the root-mean-square (\textsc{RMS}) scaled version of Parseval's theorem.


\subsection{Detecting power excess}
\label{sec:power}
We searched the power spectra of all observed stars in C2--3 for indications of seismic excess power; for the C1 cohort we adopted the four targets with clearly detected oscillations from \citet[][]{2015PASP..127.1038C}.
\begin{figure}
\centering
\includegraphics[scale=0.45]{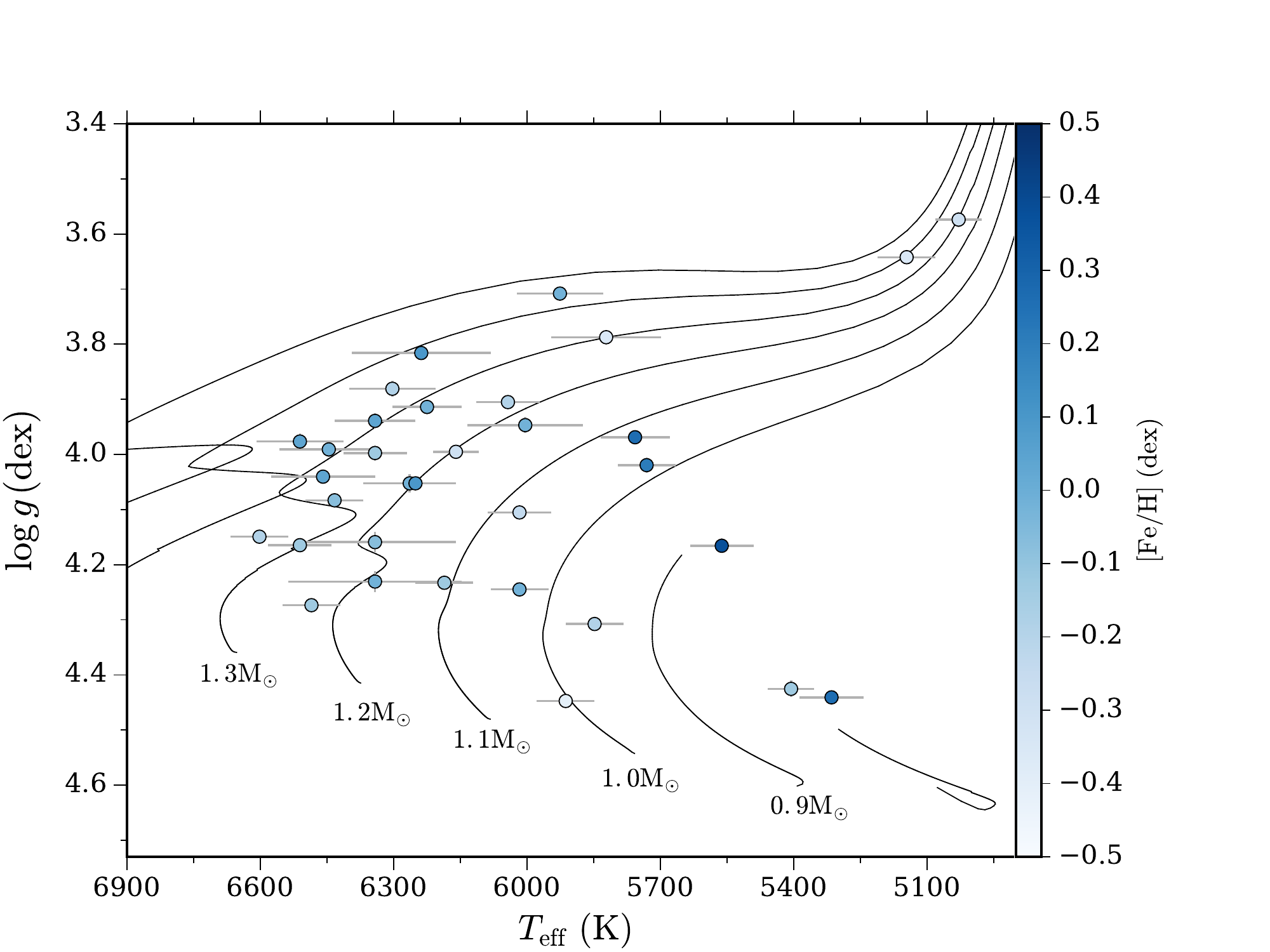}
\caption{Kiel-diagram of sample stars with detected seismic excess power, using the grid-based model results from \sref{sec:model} (see \tref{tab:model_values}). Stellar evolutionary tracks are calculated using GARSTEC adopting $\rm [Fe/H]=0$.}
\label{fig:HRdia_grid}
\end{figure}
We found 5 solid detections for C2 targets --- this yield of ${\sim}15\%$ should be seen in the context of a very crowded field in C2 with a pointing near the Galactic centre, which naturally increases the noise from aperture photometry. Also, data for both C1 and C2 were taken before the improvement in K2 SC operations, which took effect from C3 onwards.
Six additional targets from C2 did show clear oscillations, but of a classical and coherent nature rather than the stochastic nature of solar-like oscillations \citep[][]{2010aste.book.....A}. These are all seemingly members of the ``Upper Sco/Assoc. II Sco/Ass Sco OB 2-2'' associations \citep[][]{2000MNRAS.313...43H,2011MNRAS.416.3108R,2012ApJ...758...31L}, and therefore possibly pre-main-sequence oscillators \citep[][]{2014Sci...345..550Z}.
The yield from C3 was, with $24$ detections, high ($73\%$), because of the improved pointing stability.
Based on the detections in C1, \citet[][]{2015PASP..127.1038C} predicted that solar-like oscillations should be detected in C3 up to \numax values of ${\sim}2500\, \rm\mu Hz$ based on an anticipated shot-noise level three times that of the nominal \kp mission from C3 onwards \citep[][]{2016van.cleve.pasp}.
Of the targets with detected power excess in C3 there are three with \numax above the solar value of ${\sim}3090\, \rm \mu Hz$. This shows that the noise levels are better than expected (see \sref{sec:fut}), and that detections of solar-like oscillations in main-sequence and sub-giant stars should be readily achievable with K2 SC observations from C3 onwards. 
\begin{figure}
\centering
\includegraphics[scale=0.36]{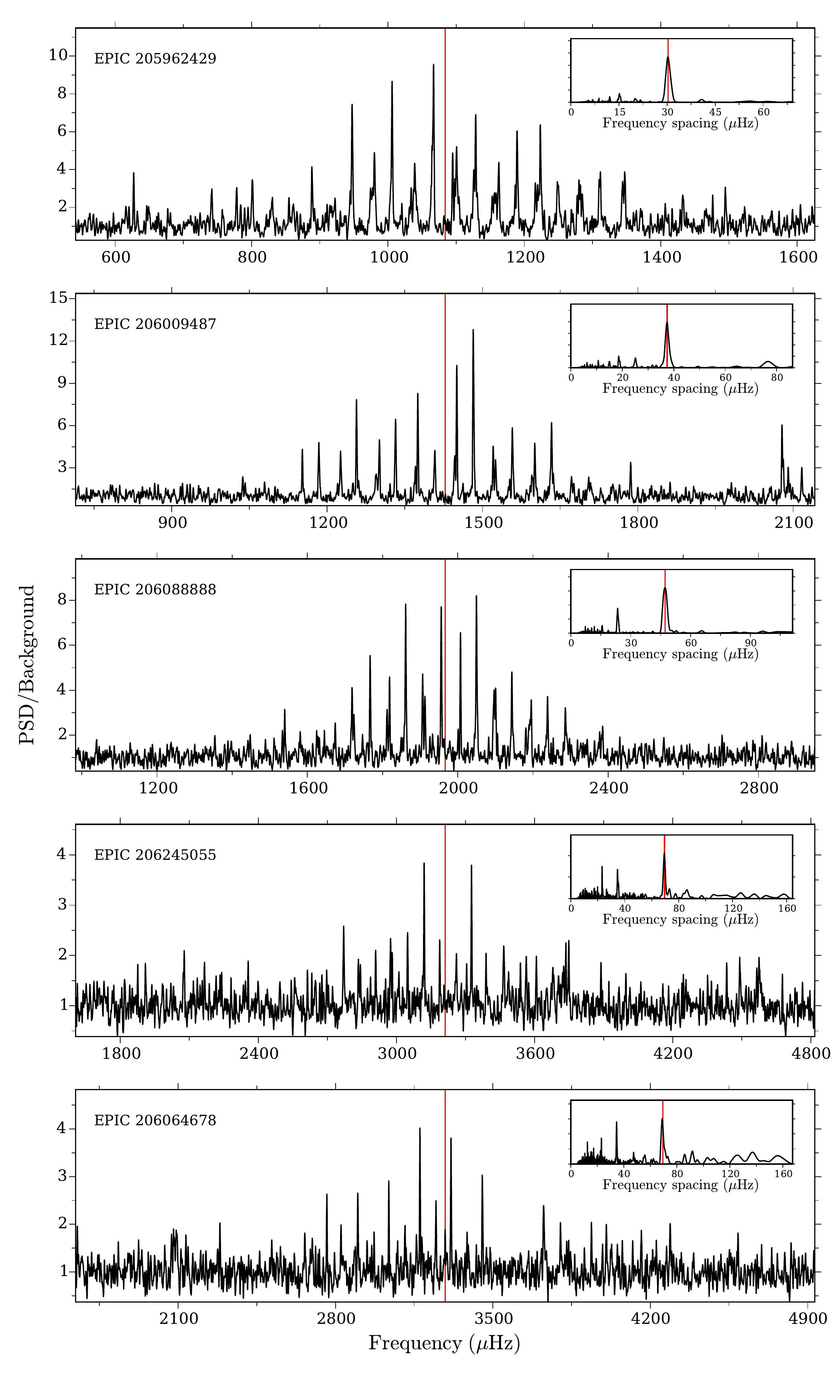}
\caption{Background-corrected power spectra for five of the targets for which individual frequencies were extracted, arranged (downwards) in order of increasing \numax and each smoothed with a $\dnu/25\,\, \rm\mu Hz$ Epanechnikov filter \citep[][]{16463184,hastie2009elements} --- two of these have \numax values above the solar value. The inserts show the power-of-power spectrum ($\rm PS\otimes PS$), where the most prominent peak (marked with a red vertical line) indicated the value of $\Delta\nu/2$. The vertical red lines in the main figure window indicate the estimated \numax values (see \sref{sec:param}). }
\label{fig:ps}
\end{figure}
In \fref{fig:HRdia_grid} we show a Kiel (spectroscopic Hertzsprung-Russell) diagram \citep[][]{2014A&A...564A..52L} of the stars with detected oscillations. See \tref{tab:parameter_values} for further information on the targets.
In \fref{fig:ps} we show the frequency-power spectra for five of the targets where individual frequencies were extracted, including two of the targets with \numax higher than the Sun. In \sref{sec:noise} we return to the detectability of seismic power for the targets.
We note that all analysis has been done using SC data which are affected by an incorrect pixel calibration\footnote{see \url{http://archive.stsci.edu/kepler/KSCI-19080-002.pdf} for further details.}, a flaw that was recently discovered. The effect of the incorrect calibration is largely dependent on the proximity of nearby bright targets that could contribute signal to a given target star. The time variation of such a signal should correlate with the apparent movement on the CCD, hence should be corrected for in our light curve preparation. Indeed, in comparing the power spectra from the recalibrated SC C3 data that was recently released with those used in this study we found no significant improvement in the shot-noise properties. We do not expect an impact on the measured seismic parameters in the current study from adopting recalibrated data.


\subsection{Atmospheric parameters}

Two sets of atmospheric parameters were obtained, one from spectroscopy and one from the InfraRed Flux Method \citep[IRFM; see][]{2014ApJ...787..110C}. Below we go through these in turn.

\subsubsection{Spectroscopic estimates}
We obtained ground-based spectroscopic data for the targets in C1 and 3 from the Tillinghast Reflector Echelle Spectrograph \citep[TRES;][]{2007RMxAC..28..129S, furesz_phd} on the 1.5-m Tillinghast telescope at the F.~L.~Whipple Observatory. The Stellar Parameter Classification pipeline \citep[SPC;][]{2012Natur.486..375B} was used to derive atmospheric parameters.
Because of well-known degeneracies between spectroscopic estimates for \teff, $\log g$, and \feh \citep[][]{2005MSAIS...8..130S,2012ApJ...757..161T} the SPC results were refined by an iterative procedure \citep[][]{2012MNRAS.423..122B}. Here the initial \teff was used together with the measured \numax to estimate the seismic $\log g$ as
\begin{equation}\label{eq:logg}
g \simeq g_{\sun} \left( \frac{\numax}{\nu_{\rm max,\sun}} \right) \left( \frac{T_{\rm eff}}{T_{\rm eff, \sun}} \right)^{1/2}\, ,
\end{equation}
using $\nu_{\rm max,\sun} = 3090\pm30\, \rm \mu Hz$, $T_{\rm eff, \sun} = 5777$ K, and $g_{\sun}=27402\, \rm cm\, s^{-2}$ \citep[][]{1991ApJ...368..599B,1995A&A...293...87K,2011ApJ...743..143H,2014ApJS..210....1C}.
The SPC analysis was then re-run with $\log g$ fixed to this seismic value --- convergence was generally obtained after a single iteration.
We added systematic uncertainties of $59$ K and $0.062$ dex in quadrature to the \teff and \feh estimates from SPC \citep[see][]{2012ApJ...757..161T}.

\subsubsection{InfraRed Flux Method}
\label{sec:irfm}

We also estimated \teff by means of the IRFM using broadband $JHK_s$ photometry from the Two Micron All Sky Survey \citep[2MASS;][]{2003yCat.2246....0C,2006AJ....131.1163S}. For targets with detected seismic power excess the IRFM solution was iterated against the asteroseismic surface gravity (see Equation~\ref{eq:logg}).
In the iteration we  first made a MCMC fit of the \teff - $\log g$ solution from the IRFM evaluated at a range of $\log g$ values. Here the uncertainty on \teff for a given $\log g$ was obtained from the scatter between the $JHK_s$ bands. Starting the iteration from the spectroscopic \teff, we evaluate the IRFM \teff from the seismic $\log g$ (Equation~\ref{eq:logg}) which is then fed back in the iteration; the iteration is done in a Monte Carlo manner where we draw randomly from the posteriors of the \teff - $\log g$ fit and the solar and stellar \numax-values.
As a valuable by-product of the IRFM one obtains a measure of the stellar angular diameter, $\theta$, which was iterated in the same manner as \teff.
Interstellar reddening, $E(B-V)$, was included in the IRFM solution by adopting values from the Geneva-Copenhagen Survey \citep[GCS;][]{2004A&A...418..989N,2011A&A...530A.138C} if available. Otherwise, field-average values of $E(B-V)=0$ for C1 and 3, and $E(B-V)=0.02$ for C2 (which is pointed close towards the Galactic centre) were adopted. The averages were obtained from the reddening of field stars in the GCS within ${\sim}10^{\circ}$ of the centres of the C1, C2 and C3 pointings and beyond 40 pc in distance; for stars closer than this a zero reddening was adopted in the GCS as appropriate to the near Solar Neighbourhood \citep[see][]{2007A&A...475..519H}. The effect of reddening on the IRFM \teff typically amounts to ${\sim}50$ K per $0.01$ mag excess \citep[see][]{2010A&A...512A..54C}.
To evaluate the uncertainty on \teff due to the uncertainty in reddening we tried adopting $E(B-V)$ values from the 3D dust map by \citet[][]{2015ApJ...810...25G}\footnote{using the Python API available at \url{http://argonaut.skymaps.info/}}, derived from stars in the Pan-STARRS 1 survey. Here $E(B-V)$ was taken as the interpolated median reddening solution at the (parallax) distance of the given star. Unfortunately, most stars lie within the minimum distance deemed reliable by \citet[][]{2015ApJ...810...25G} on the grounds of low numbers of stars, and the adopted reddening values should therefore be treated with caution. As a systematic uncertainty we added in quadrature the differences in \teff and $\theta$ between this solution and the one using field average reddenings to the IRFM uncertainties. An additional systematic uncertainty from a Monte Carlo run assigning photometry errors, $\rm [Fe/H]$ uncertainties of $0.2$ dex, and reddening uncertainties of $0.01$ mag was also added. We finally added zero-point uncertainties of $20$ K in \teff and $0.7\%$ on $\theta$.

Three of the C2 targets (204356572, 204550630, and 204926239) have uncertainties in the IFRM determination of \teff in excess of $1000$ K, because these targets are possibly subject to high levels of reddening. The targets are indeed found to lie in close proximity to the centre of the Rho Ophiuchi cloud complex.
For one of these targets, EPIC 204926239, we have followed a different approach than outlined above and instead adopted the IRFM \teff and $\theta$ solution that uses the $E(B-V)$ from \citet[][]{2015ApJ...810...25G}. The resulting \teff agrees with that obtained by \citet[][]{2012ApJ...746..154P}. In \sref{sec:239} we discuss the results for this target further.

\begin{table*} 
\renewcommand{\arraystretch}{1}
\centering
\caption{
Parameters for the 33 target under study. For targets where the EPIC number is in boldface individual frequencies have been extracted. ``Cam.'' gives the K2 campaign; ``Kp'' gives the \kp magnitude \citep[][]{2011AJ....142..112B,2015arXiv151202643H}; ``HIP. ID'' gives the \textit{Hipparcos} identifier of the target; ``$\pi$'' gives the \citet[][]{2007A&A...474..653V} \textit{Hipparcos} parallax in milli-arc-seconds (mas); ``$\theta$'' is the stellar angular diameter from the IRFM in mas; ``LOS'' gives the line-of-sight velocity from the CfA TRES observations, corrected by $-0.61\, \rm km/s$. } 
\label{tab:parameter_values}
\mysize
\begin{tabular}{@{\extracolsep{1.5pt}}l@{\hskip -0.05cm}ccccccccccccc@{}} %
\toprule 
\multicolumn{3}{c}{K2} & \multicolumn{2}{c}{Hipparcos}  &  \multicolumn{2}{c}{Seismic}  & \multicolumn{2}{c}{IRFM} & \multicolumn{5}{c}{SPC} \\ 
\cline{1-3}\cline{4-5}\cline{6-7}\cline{8-9}\cline{10-14}\\[-0.5em]
EPIC & Cam. & Kp    & HIP. ID & $\pi$  & $\nu_{\rm max}$ & $\Delta\nu$    & $\theta$ & \teff & \teff & [Fe/H]            & $\log g$                    & $v\sin\, i_{\star}$ & LOS \\ 
     &      & (mag) &     & (mas)  & ($\rm \mu Hz$)  & ($\rm \mu Hz$) & (mas)    & (K)           & ($\pm 77$ K)  & ($\pm 0.10$ dex)  & ($\rm cgs$; $\pm 0.10$ dex) & ($\pm 0.5 \, \rm km\ s^{-1}$) & ($\rm km\ s^{-1}$)\\
\midrule
$201367296$               & 1 & $7.439$ & $58093$  & $16.68 \pm 0.88$  &  $1176 \pm 58$  &  $64.8 \pm 2.0$  &  $0.262 \pm 0.006$   &  $5746 \pm 70$    &  $5740$   &  $0.231 $  &  $4.015$   &  $3.39 $ & $19.66  \pm 0.02$ \\
$201367904$               & 1 & $8.440$ & $58191$  & $8.39  \pm 1.02$  &  $890  \pm 46$  &  $51.6 \pm 1.5$  &  $0.141 \pm 0.003$   &  $6241 \pm 77$    &  $6270$   &  $0.048 $  &  $3.910$   &  $10.12$ & $3.47   \pm 0.03$ \\
$201820830$               & 1 & $8.766$ & $55778$  & $8.33  \pm 0.71$  &  $1196 \pm 72$  &  $67.2 \pm 2.0$  &  $0.137 \pm 0.003$   &  $6276 \pm 117$   &  $6456$   &  $0.001 $  &  $4.048$   &  $12.26$ & $7.83   \pm 0.12$ \\
$201860743$               & 1 & $7.797$ & $57676$  & $12.26 \pm 0.97$  &  $1000 \pm 46$  &  $56.1 \pm 1.8$  &  $0.215 \pm 0.005$   &  $5998 \pm 134$   &  $5930$   &  $-0.036$  &  $3.951$   &  $4.46 $ & $-1.40  \pm 0.06$ \\
$\boldsymbol{204506926}$  & 2 & $8.676$ & $81413$  & $9.76  \pm 1.08$  &  $1702 \pm 70$  &  $84.4 \pm 2.5$  &  $0.163 \pm 0.003$   &  $5570 \pm 70$    &  $5711$   &  $0.390 $  &  $4.177$   &  $3.21 $ & $9.42   \pm 0.05$ \\
$204356572$               & 2 & $7.937$ & $80374$  & $11.53 \pm 0.96$  &  $1600 \pm 95$  &  $79.6 \pm 2.2$  &  $0.174 \pm 0.014$   &  $6349 \pm 1809$  &  $6324$   &  $-0.076$  &  $4.168$   &  $6.20 $ & $5.38   \pm 0.03$ \\
$204550630$               & 2 & $8.775$ & $81235$  & $9.920 \pm 1.08$  &  $1885 \pm 130$ &  $90.4 \pm 2.3$  &  $0.121 \pm 0.010$   &  $6374 \pm 1695$  &  $6268$   &  $-0.031$  &  $4.241$   &  $10.09$ & $-31.57 \pm 0.03$ \\
$204624076$               & 2 & $8.506$ & $80756$  & $9.530 \pm 1.10$  &  $1236 \pm 80$  &  $59.4 \pm 1.4$  &  $0.158 \pm 0.003$   &  $6336 \pm 77$    &  $6370$   &  $-0.156$  &  $4.061$   &  $10.94$ & $-13.34 \pm 0.06$ \\
$204926239^c$             & 2 & $9.039$ & $79606$  & $7.720 \pm 1.42$  &  $747  \pm 31$  &  $41.5 \pm 1.1$  &  $0.161 \pm 0.004^b$ &  $6238 \pm 300^b$ &  $6225$   &  $0.082 $  &  $3.841$   &  $24.51$ & $-23.37 \pm 0.16$ \\
$205917956$               & 3 & $8.246$ & $111312$ & $13.83 \pm 0.62$  &  $527  \pm 16$  &  $33.6 \pm 1.6$  &  $0.349 \pm 0.007$   &  $5163 \pm 70$    &  $5204$   &  $-0.285$  &  $3.669$   &  $1.54 $ & $31.11  \pm 0.09$ \\
$\boldsymbol{205962429}$  & 3 & $8.963$ & $110537$ & $6.96  \pm 0.96$  &  $1084 \pm 36$  &  $60.6 \pm 1.9$  &  $0.119 \pm 0.003$   &  $6216 \pm 70$    &  $6109$   &  $-0.221$  &  $3.988$   &  $7.10 $ & $32.45  \pm 0.10$ \\                                                                                                 
$205967173$               & 3 & $6.333$ & $109822$ & $26.26 \pm 0.53$  &  $452  \pm 14$  &  $31.5 \pm 4.3$  &  $0.597 \pm 0.015$   &  $5016 \pm 70$    &  $4981$   &  $-0.261$  &  $3.578$   &  $0.95 $ & $10.48  \pm 0.02$ \\ 
$205974115^a$             & 3 & $8.179$ & $110689$ & $11.33 \pm 0.98$  &  $1914 \pm 63$  &  $89.6 \pm 2.4$  &  $0.161 \pm 0.003$   &  $6224 \pm 70$    &  $6143$   &  $-0.044$  &  $4.243$   &  $12.55$ & $ -3.61 \pm 0.05$ \\                                                                                                 
$205979004$               & 3 & $8.394$ & $110454$ & $6.73  \pm 1.16$  &  $692  \pm 22$  &  $42.8 \pm 1.7$  &  $0.175 \pm 0.004$   &  $5831 \pm 120$   &  $5601$   &  $-0.346$  &  $3.781$   &  $3.39 $ & $-29.24 \pm 0.06$ \\ 
$205995584$               & 3 & $6.972$ & $110518$ & $15.26 \pm 0.62$  &  $1341 \pm 59$  &  $67.2 \pm 1.8$  &  $0.264 \pm 0.005$   &  $6451 \pm 70$    &  $6443$   &  $-0.034$  &  $4.097$   &  $18.24$ & $ 11.82 \pm 0.09$ \\ 
$\boldsymbol{206009487}$  & 3 & $7.708$ & $111892$ & $16.10 \pm 1.68$  &  $1428 \pm 46$  &  $74.9 \pm 2.4$  &  $0.215 \pm 0.004$   &  $6030 \pm 70$    &  $5924$   &  $-0.213$  &  $4.108$   &  $3.83 $ & $-17.42 \pm 0.05$ \\
$\boldsymbol{206064678}$  & 3 & $8.831$ & $109672$ & $15.53 \pm 1.31$  &  $3288 \pm 141$ &  $138.9\pm 3.6$  &  $0.176 \pm 0.003$   &  $5326 \pm 70$    &  $5580$   &  $0.309 $  &  $4.457$   &  $1.62 $ & $19.13  \pm 0.05$ \\ 
$206064711$               & 3 & $7.308$ & $108692$ & $14.62 \pm 0.78$  &  $1563 \pm 54$  &  $74.2 \pm 1.8$  &  $0.216 \pm 0.004$   &  $6612 \pm 70$    &  $6557$   &  $-0.156$  &  $4.165$   &  $14.92$ & $ 6.53  \pm 0.07$ \\ 
$206070413$               & 3 & $8.47$  & $111534$ & $7.64  \pm 1.09$  &  $1550 \pm 76$  &  $79.2 \pm 2.1$  &  $0.129 \pm 0.003$   &  $6528 \pm 73$    &  $6458$   &  $-0.071$  &  $4.161$   &  $20.75$ & $6.83   \pm 0.10$ \\
$\boldsymbol{206078331}$  & 3 & $7.033$ & $108468$ & $29.93 \pm 0.74$  &  $2253 \pm 76$  &  $111.2\pm 3.1$  &  $0.308 \pm 0.006$   &  $5861 \pm 70$    &  $5818$   &  $-0.144$  &  $4.303$   &  $2.59 $ & $3.88   \pm 0.02$ \\ 
$\boldsymbol{206088888}$  & 3 & $8.718$ & $111376$ & $10.35 \pm 1.38$  &  $1967 \pm 64$  &  $94.4 \pm 2.7$  &  $0.132 \pm 0.003$   &  $6029 \pm 70$    &  $5962$   &  $-0.036$  &  $4.248$   &  $3.74 $ & $45.83  \pm 0.04$ \\ 
$206094605$               & 3 & $8.638$ & $110065$ & $5.58  \pm 0.91$  &  $1027 \pm 50$  &  $55.7 \pm 1.4$  &  $0.119 \pm 0.003$   &  $6513 \pm 104$   &  $6546$   &  $0.013 $  &  $3.985$   &  $27.42$ & $15.44  \pm 0.14$ \\ 
$206107253$               & 3 & $8.439$ & $110217$ & $11.37 \pm 1.00$  &  $2053 \pm 64$  &  $95.2 \pm 2.4$  &  $0.130 \pm 0.003$   &  $6479 \pm 70$    &  $6315$   &  $-0.139$  &  $4.274$   &  $9.24 $ & $-21.59 \pm 0.07$ \\ 
$206108325^a$             & 3 & $8.294$ & $110902$ & $7.00  \pm 1.36$  &  $861  \pm 54$  &  $48.3 \pm 1.4$  &  $0.151 \pm 0.003$   &  $6319 \pm 105$   &  $6446$   &  $-0.150$  &  $3.903$   &  $22.10$ & $18.78  \pm 0.29$ \\ 
$206184719$               & 3 & $6.824$ & $111843$ & $14.56 \pm 0.81$  &  $557  \pm 19$  &  $36.7 \pm 1.6$  &  $0.331 \pm 0.007$   &  $5911 \pm 104$   &  $5855$   &  $0.029 $  &  $3.700$   &  $8.75 $ & $-25.19 \pm 0.05$ \\
$\boldsymbol{206245055}$  & 3 & $8.899$ & $111332$ & $16.17 \pm 0.96$  &  $3212 \pm 119$ &  $138.9\pm 3.5$  &  $0.145 \pm 0.003$   &  $5917 \pm 70$    &  $5846$   &  $-0.375$  &  $4.455$   &  $2.84 $ & $-33.12 \pm 0.03$ \\ 
$206289767$               & 3 & $8.820$ & $109899$ & $6.07  \pm 1.13$  &  $972  \pm 42$  &  $52.2 \pm 1.4$  &  $0.127 \pm 0.003$   &  $6343 \pm 87$    &  $6367$   &  $0.058 $  &  $3.953$   &  $17.13$ & $10.09  \pm 0.05$ \\ 
$206368174$               & 3 & $8.276$ & $110002$ & $10.56 \pm 1.03$  &  $1252 \pm 42$  &  $63.9 \pm 1.8$  &  $0.150 \pm 0.003$   &  $6246 \pm 70$    &  $6201$   &  $0.079 $  &  $4.055$   &  $5.43 $ & $1.40   \pm 0.04$ \\ 
$206371648^a$             & 3 & $8.605$ & $109951$ & $16.09 \pm 1.07$  &  $3047 \pm 199$ &  $137.0\pm 4.8$  &  $0.189 \pm 0.004$   &  $5327 \pm 70$    &  $5535$   &  $-0.249$  &  $4.424$   &  $4.88 $ & $-23.27 \pm 0.11$ \\
$\boldsymbol{206445085}$  & 3 & $8.462$ & $109836$ & $12.14 \pm 0.76$  &  $1060 \pm 34$  &  $58.3 \pm 1.9$  &  $0.236 \pm 0.005$   &  $5758 \pm 80$    &  $5717$   &  $0.271 $  &  $3.976$   &  $3.70 $ & $8.41   \pm 0.04$ \\ 
$206453540$               & 3 & $8.582$ & $109783$ & $9.03  \pm 0.71$  &  $1218 \pm 46$  &  $61.3 \pm 1.4$  &  $0.176 \pm 0.004$   &  $6494 \pm 135$   &  $6543$   &  $0.020 $  &  $4.061$   &  $24.89$ & $-12.74 \pm 0.10$ \\
$206189649^a$             & 3 & $8.537$ & $110974$ & $10.10 \pm 1.23$  &  $895  \pm 31$  &  $52.2 \pm 1.7$  &  $0.147 \pm 0.003$   &  $6049 \pm 70$    &  $6088$   &  $-0.147$  &  $3.910$   &  $7.71 $ & $15.37  \pm 0.07$ \\
$206201061$               & 3 & $8.749$ & $110077$ & $6.45  \pm 1.00$  &  $1040 \pm 39$  &  $58.0 \pm 1.6$  &  $0.116 \pm 0.002$   &  $6471 \pm 115$   &  $6605$   &  $0.029 $  &  $3.991$   &  $13.77$ & $-2.95  \pm 0.23$ \\
\bottomrule
\end{tabular}
\begin{tablenotes}[normal]  
  \scriptsize 
	\item $^a$Double star in The Washington Double Star Catalog \citep[WDS;][]{1997A&AS..125..523W}; $^b$Calculated assuming an $E(B-V)$ from \citet[][]{2015ApJ...810...25G}; the obtained \teff agrees with that obtained by \citet[][]{2012ApJ...746..154P}; $^c$Potential member of the Upper Scorpius association \citep[][]{1999AJ....117..354D}.
 \end{tablenotes}
\end{table*} 


\section{Seismic parameter estimation}
\label{sec:param}

\begin{figure}
\centering
\includegraphics[width=\columnwidth]{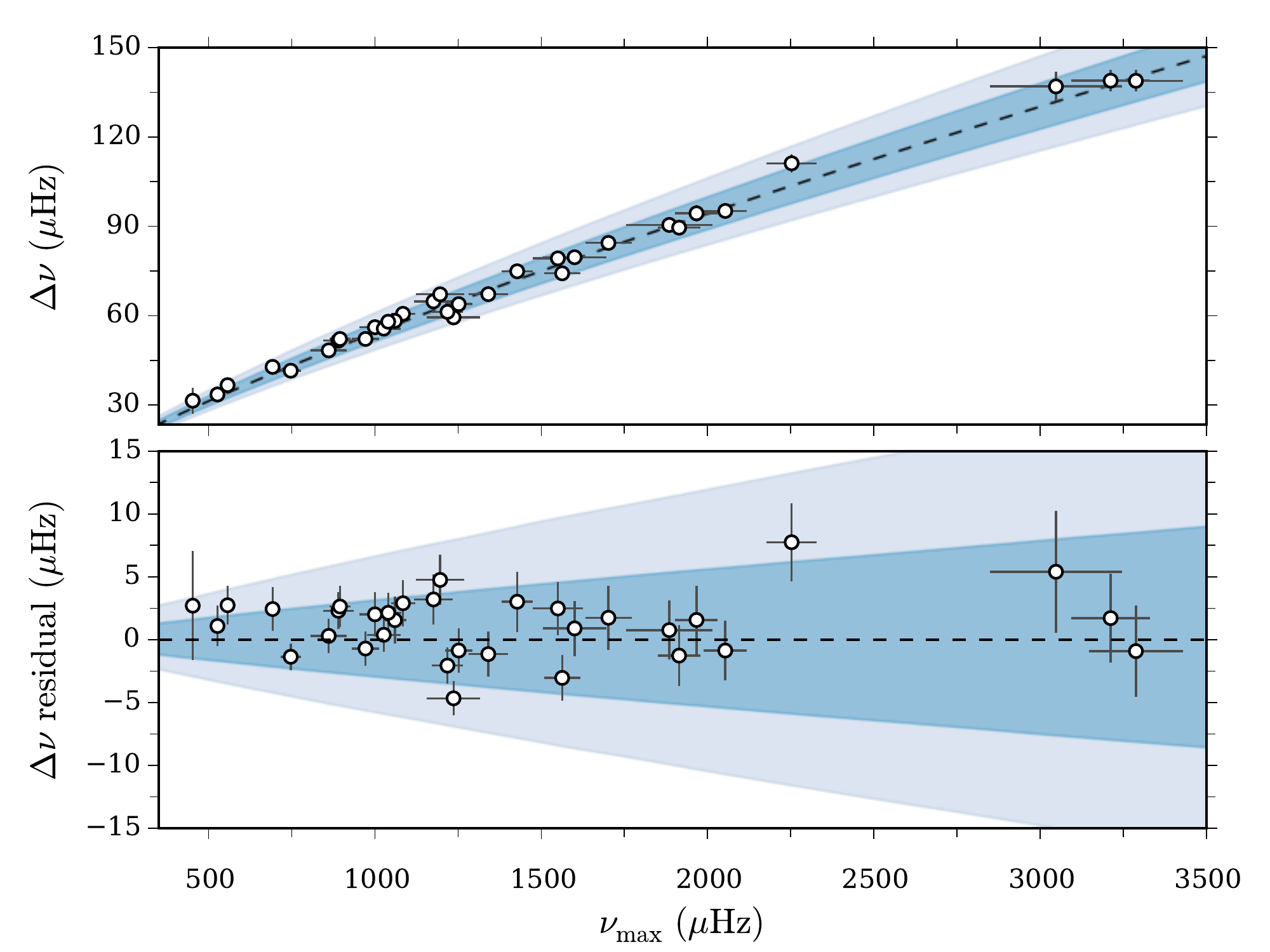}
\caption{Top: Estimated \numax against \dnu; the dashed line gives the relation by \citet[][]{2011ApJ...743..143H}, with the 1 and $2\sigma$ confidence intervals indicated by the dark and light blue shaded regions, respectively. Bottom: Residual between measured \dnu values and the scaling relation.}
\label{fig:dnu_numax}
\end{figure}

Global asteroseismic parameters \dnu and \numax were estimated for all targets with detected power excess --- for the C1 targets we adopted the values presented in \citet[][]{2015PASP..127.1038C}.
For all remaining stars the estimation of \numax was achieved by fitting the stellar noise background following \citet[][]{2014A&A...570A..54L}. We adopted a background model given by a sum of generalised Lorentzians with free exponents, time scales, and rms amplitudes \citep[see][]{1985ESASP.235..199H,2012MNRAS.421.3170K,2014A&A...570A..41K}, and a Gaussian envelope centred on \numax was adopted to account for the power excess due to the oscillation spectrum. A systematic fractional uncertainty of $3\%$ on \numax was added in quadrature following \citet[][]{2011MNRAS.415.3539V} who found this to be the average difference between different methods for estimating \numax; the median fractional uncertainty on \numax amounted to ${\sim}3.7\%$. We estimated \dnu from the peak value of the power-of-power spectrum ($\rm PS\otimes PS$) centred on $\Delta\nu/2$, where the guess of \dnu was obtained from the $\Delta\nu\propto \beta\numax^{\alpha}$ scaling by \citet[][]{2011ApJ...743..143H}. The FWHM of the $\Delta\nu/2$ peak was adopted as a conservative uncertainty on \dnu; this gives a median fractional uncertainty on \dnu of ${\sim}2.8\%$.
In \fref{fig:dnu_numax} we show the relation between the measured \numax and \dnu values, together with the \citet[][]{2011ApJ...743..143H} scaling. As seen the measured values conform to the scaling relation. See \tref{tab:parameter_values} for the extracted \numax and \dnu values.

\begin{figure}
\centering
\includegraphics[width=\columnwidth]{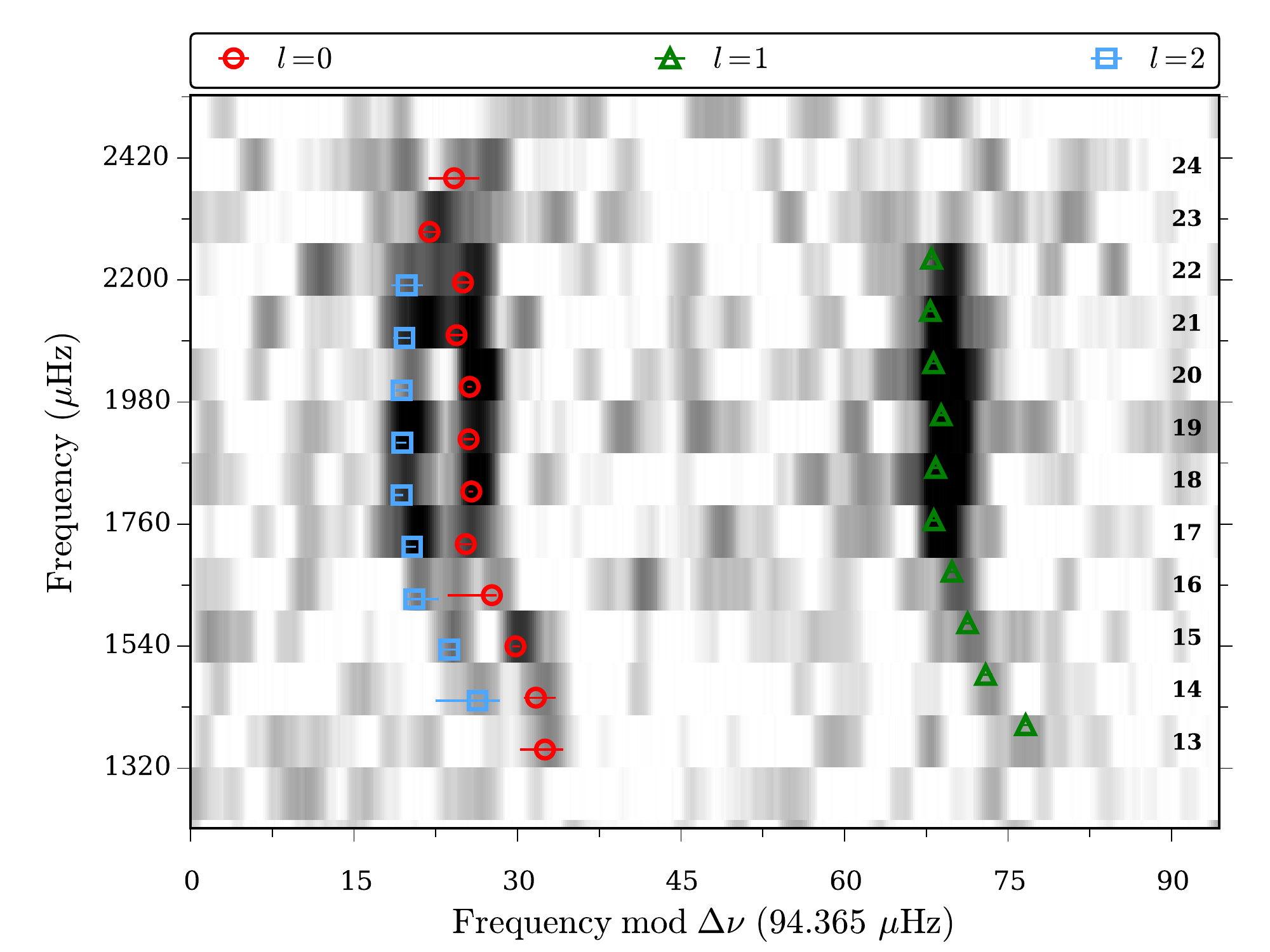} 
\caption{\'{E}chelle diagram for EPIC 206088888. The grey scale indicate the power level going from white (low) to black (high). The markers indicate the extracted mode frequencies, see the legend for the mode degree; the radial order of the $l=0$ modes is indicated by the numbers on the right-hand side of the figure.}
\label{fig:echelle}
\end{figure}

For a select number of $8$ targets we also extracted individual frequencies by peak-bagging the power spectra. More of the targets could be peak-bagged; however, due to the relatively lower signal-to-noise ratio (SNR) of these targets this is beyond the scope of the current work where we simply wish to assert the potential of SC observations with K2, rather than provide a full in-depth modelling of all SC targets. 
The peak-bagging was performed as described in \citet[][]{2014A&A...570A..54L}, that is, fitting a global model using an MCMC affine invariant ensemble sampler \citep[see][]{2013PASP..125..306F}. Mode identification in terms of angular degree was achieved using the $\epsilon$ vs. \teff relation given in \citet[][]{2011ApJ...742L...3W,2012ApJ...751L..36W}. \fref{fig:echelle} shows the \'{e}chelle diagram \citep[][]{1983SoPh...82...55G,2010CoAst.161....3B} for EPIC 206088888, for which individual frequencies could be extracted (see also \fref{fig:ps}).

\begin{figure}
\centering
\includegraphics[width=\columnwidth]{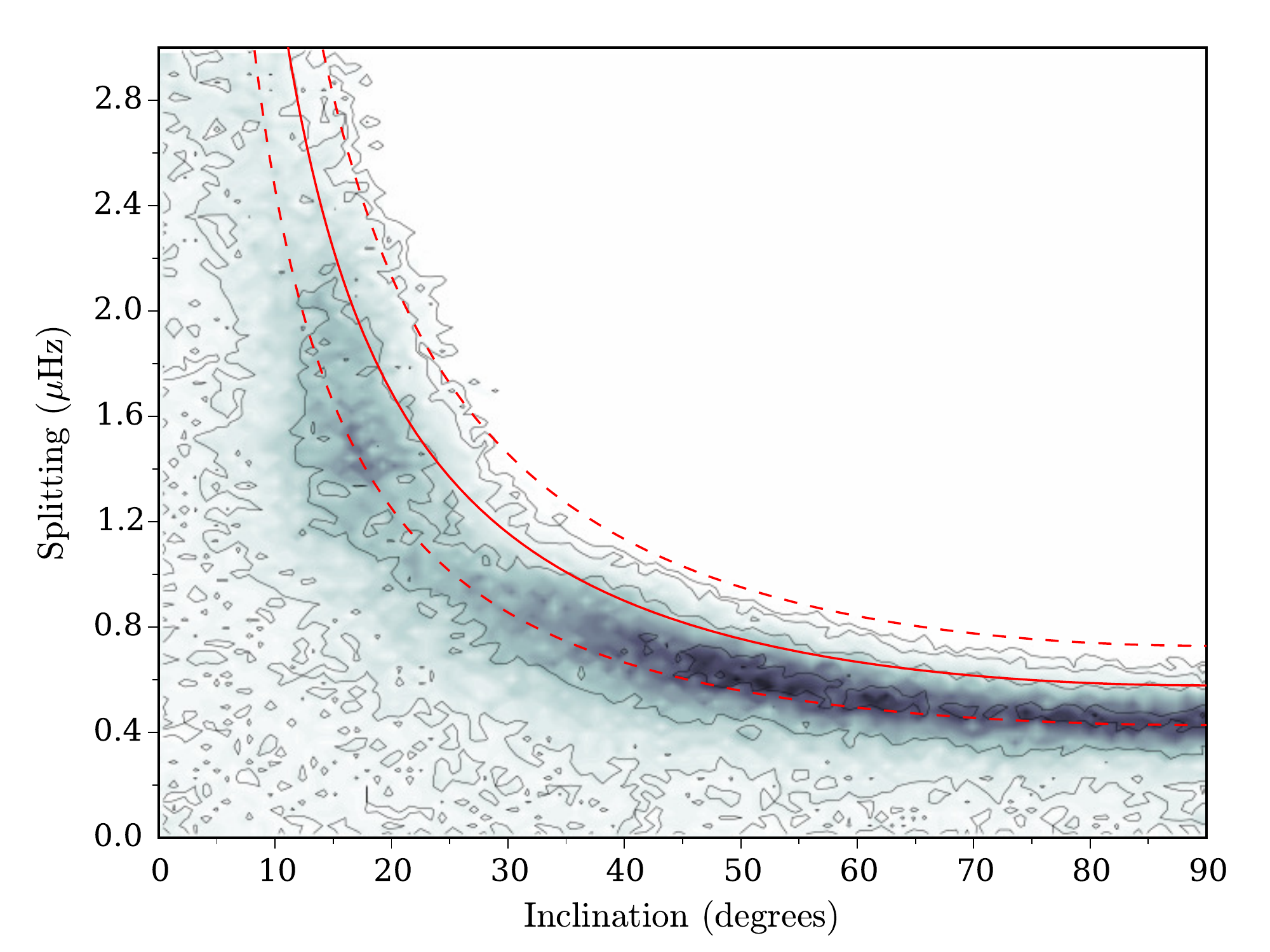}
\caption{Correlation maps from the MCMC fits of inclination versus frequency splitting for EPIC 206009487. The colour scale goes from low (white) to high posterior density (blue). The solid red line gives the splitting for a constant $v\sin\, i_{\star}$ with corresponding uncertainties (dashed lines), computed from the derived stellar radius and spectroscopic $v\sin\, i_{\star}$ (see Tables~\ref{tab:parameter_values} and \ref{tab:model_values}).}
\label{fig:bananna}
\end{figure}

From the MCMC peak-bagging one obtains posterior distributions for parameters such as the frequency splitting $\delta\nu_s$ due to rotation and the inclination angle $i_{\star}$ of the star \citep[see, \eg,][]{2013ApJ...766..101C,2013Sci...342..331H,2014A&A...570A..54L,2015MNRAS.446.2959D}. \fref{fig:bananna} shows the splitting versus inclination correlation map from the MCMC fit of EPIC 206009487. As seen, one can recover the curved ``banana-shaped'' correlation indicating a constant $\delta\nu_s \sin i_{\star}$, corresponding to a given projected rotational velocity $v\sin i_{\star}$. In the figure we have indicated the lines of equivalent constant $\delta\nu_s \sin i_{\star}$ from the measured spectroscopic $v\sin i_{\star}$ (see \tref{tab:parameter_values}) and modelled stellar radius (see \tref{tab:model_values}). In addition to EPIC 206009487 we were also able to determine projected rotation rates for EPICs 206088888 and 205962429. Thus, from ${\sim}80$ days of K2 photometry one will in some stars be in a position to asses the stellar rotation, and possibly pin down the inclination from time series estimates of the rotation period.


\section{Seismic modelling}
\label{sec:model}

\begin{figure}
\centering
\includegraphics[width=\columnwidth]{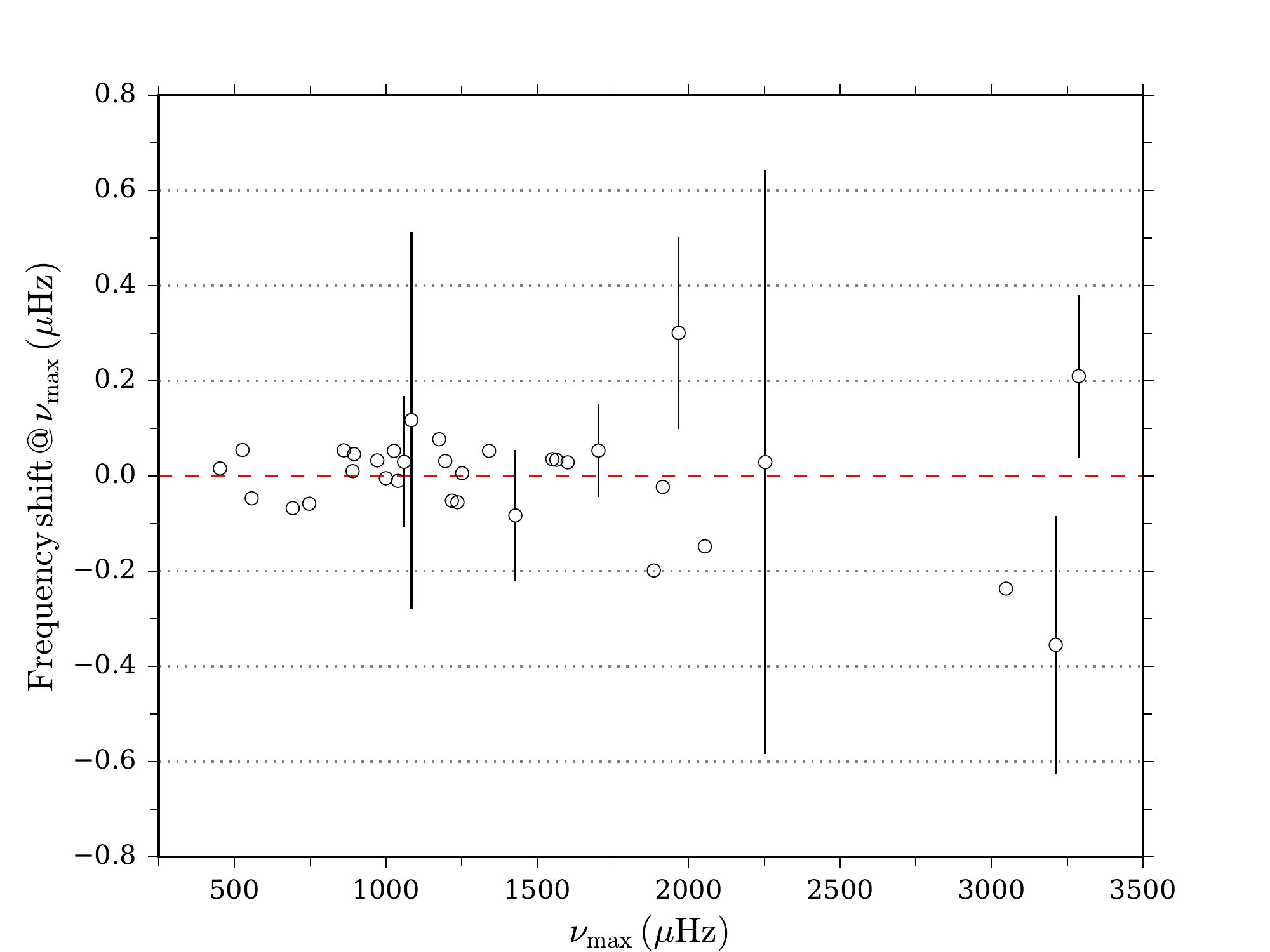}
\caption{Mode frequency shift at \numax from the line-of-sight velocity of the individual targets, obtained from the SPC data. For targets with individual frequencies from peak-bagging we have indicated the minimum frequency uncertainty of the five radial modes nearest \numax. For several of the targets the frequency shift exceeds the uncertainty on individual mode frequencies.}
\label{fig:los}
\end{figure}

In the modelling described below results were derived using the \teff from both the SPC and the IRFM; in all cases the metallicity from the SPC was adopted.
Before any modelling of individual frequencies we corrected the frequencies for the line-of-sight (LOS) velocity of the targets following \citet[][]{2014MNRAS.445L..94D}; for modelling using frequency ratios this correction is insignificant. We obtained the LOS velocities from the Mg b order in the TRES observations, and corrected by $-0.61\, \rm km/s$ to put the velocities onto the IAU system (see \tref{tab:parameter_values}).  Most of this correction is due the fact that the CfA library of synthetic spectra does not include the gravitational redshift of the Sun. 
For three of the five peak-bagged targets the correction was at the level of the frequency uncertainties (see \fref{fig:los}). Note that the frequency shift scales linearly with frequency, so modes above \numax will be shifted more than lower-frequency modes. 

Average seismic parameters were modelled using two pipelines: (1) The BAyesian STellar Algorithm \citep[BASTA;][]{2015MNRAS.452.2127S} using evolution models calculated with the Garching Stellar Evolution Code \citep[GARSTEC;][]{2008Ap&SS.316...99W} and frequencies computed with the Aarhus adiabatic oscillation package \citep[ADIPLS;][]{2008Ap&SS.316..113C}. Besides the atmospheric observables \teff and \feh, BASTA uses the seismic quantities \dnu and \numax with the model \dnu computed from individual frequencies and \numax computed using the usual scaling relation ($\nu_{\rm max}\propto g/\sqrt{T_{\rm eff}}$); (2) the Yale-Birmingham code \citep[YB; ][]{2010ApJ...710.1596B, 2012ApJ...746...76B,2011ApJ...730...63G}, which utilises three different grids with models from the Dartmouth group \citep[][]{2008ApJS..178...89D}, the Yonsei-Yale (YY) isochrones \citep[][]{2004ApJS..155..667D}, and YREC2 as described by \citet[][]{2012ApJ...746...76B}. In YB model values of \dnu were computed using the simple scaling relation between $\Delta\nu$ and density (\ie, $\Delta\nu\propto\sqrt{M/R^3}$), with a correction applied to $\Delta\nu$ following \citet[][]{2011ApJ...742L...3W} to rectify the deviations of $\Delta\nu$ from the pure scaling. YB similarly use the scaling relation for \numax. 
For solar reference values we adopted $\Delta\nu_{\rm\sun} = 135.1\pm0.1\, \rm \mu Hz$, $\nu_{\rm max,\sun} = 3090\pm30\, \rm \mu Hz$, and $T_{\rm eff, \sun} = 5777$ K \citep[][]{2011ApJ...743..143H}.


In the modelling using individual frequencies three pipelines were used: (1) the BASTA, where the frequency separation ratios $r_{010}$ and $r_{02}$ are used rather than individual frequencies directly \citep[see][]{2003A&A...411..215R,2011A&A...529A..63S,2013ApJ...769..141S} --- we shall refer to this as BASTA2 to distinguish it from the use to BASTA with average seismic parameters; (2) the ASTEC Fitting method (ASTFIT) using evolutionary models from the Aarhus STellar Evolution Code \citep[ASTEC][]{2008Ap&SS.316...13C} and frequencies from ADIPLS; (3) the Yale-Monte Carlo Method (YMCM) with evolutionary models from the Yale Stellar Evolution Code \citep[YREC;][]{2008Ap&SS.316...31D} and frequencies from the code described by \citet[][]{1994A&AS..107..421A}.
Further details on the different codes and grids are given by \citet[][]{2015MNRAS.452.2127S} and \citet[][]{2014ApJS..210....1C}.


\subsection{Comparison of model results}
\label{sec:comp}
\begin{figure}
\centering
\includegraphics[width=\columnwidth]{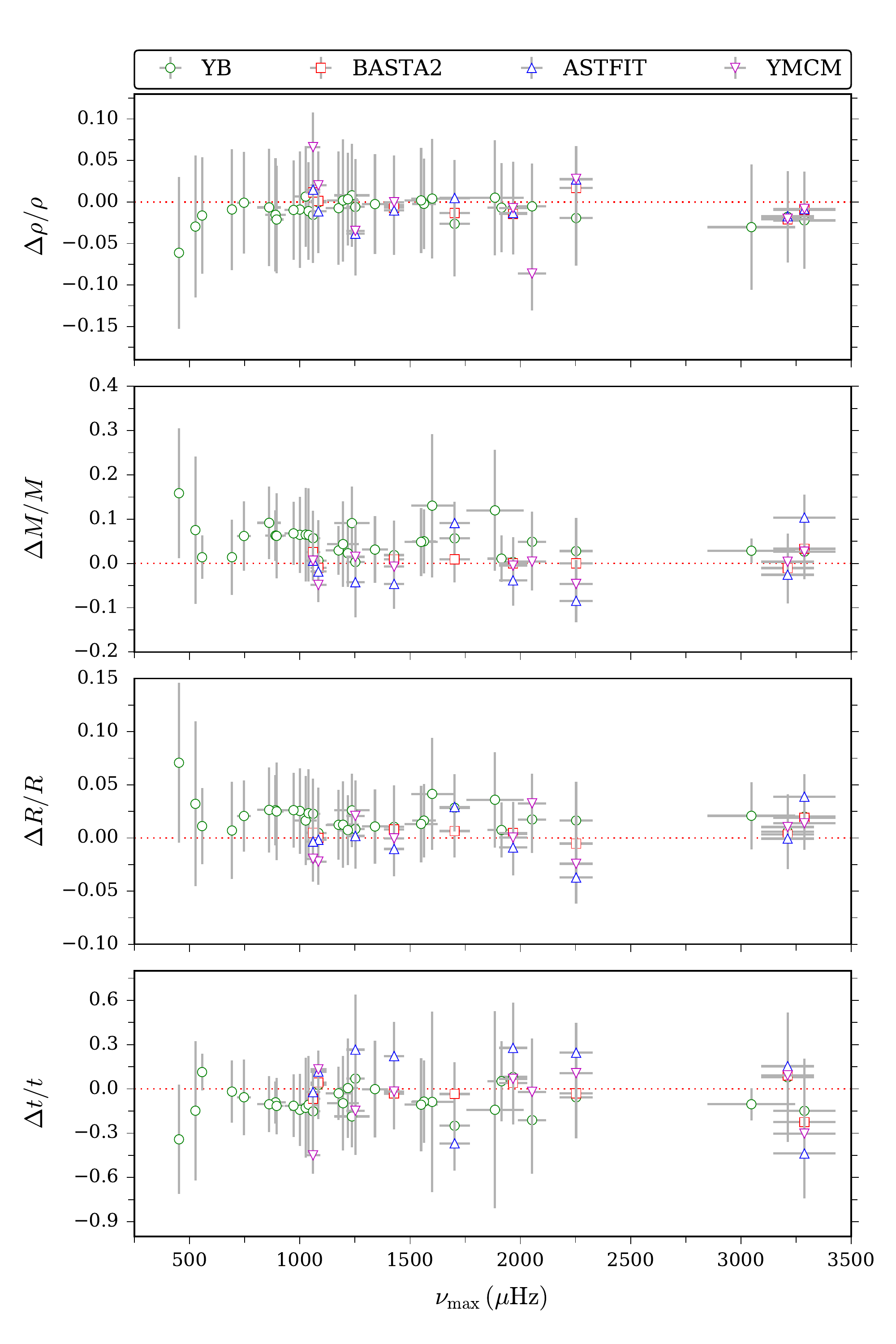}
\caption{Fractional differences between results from different modelling pipelines, or the use of either average seismic parameters or individual frequencies, against \numax. The comparison is relative to results from BASTA (as (OTHER-BASTA)/BASTA, see legend for ``OTHER'' pipeline) using average seismic parameters and inputs from the IRFM (\tref{tab:model_values}). BASTA2 indicates model results obtained using ratios from individual frequencies.}
\label{fig:model_comp}
\end{figure}
\begin{figure}
\centering
\includegraphics[width=\columnwidth]{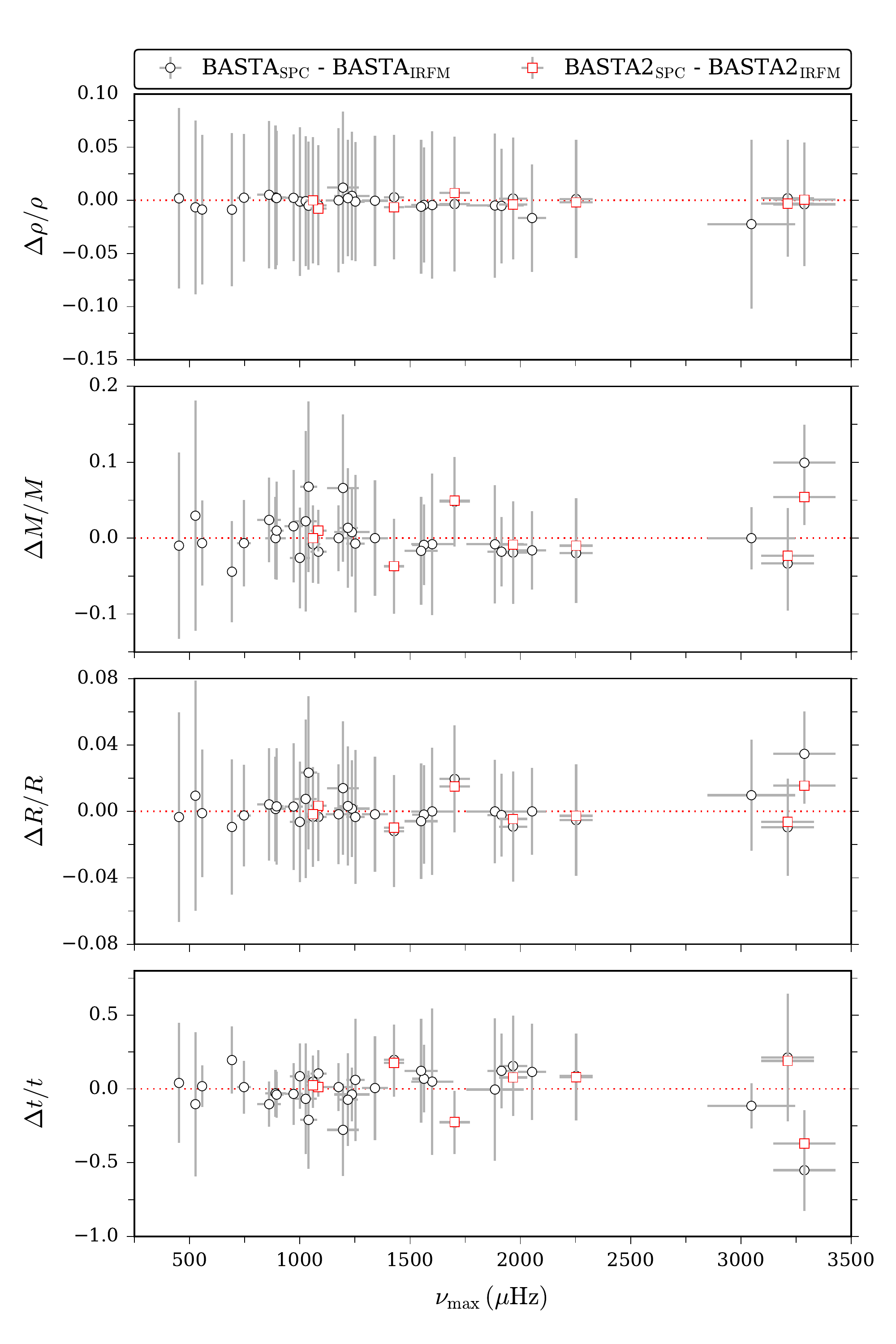}
\caption{Fractional differences between BASTA results from different spectroscopic inputs (see Tables~\ref{tab:model_values} and \ref{tab:model_values2}), against \numax. BASTA2 indicates model results obtained using ratios from individual frequencies.}
\label{fig:model_comp2}
\end{figure}
In \fref{fig:model_comp} we show the comparison between modelling results from different pipelines, with BASTA results from average seismic parameters and IRFM inputs as the reference. Overall we see a very good agreement in derived parameters, with only a few examples of differences exceeding $1\sigma$. In terms of uncertainties we obtained from the BASTA grid-based modelling formal median fractional uncertainties of $4.3\%$ in density, $4.5\%$ in mass, $2.4\%$ in radius, and $17.5\%$ in age; from the detailed modelling we obtained $1.4\%$ in density, $3.5\%$ in mass, $1.2\%$ in radius, and $16\%$ in age --- the sample size here was notably smaller, so these median values are statistically less secure. Concerning the use of different input parameters we obtain for the BASTA grid-based results absolute median fractional differences of $0.7\%$ in mass, $0.1\%$ in radius, $1.2\%$ in age, and $0.1\%$ in density between using IRFM vs. SPC input for \teff. The comparison is shown in \fref{fig:model_comp2}. 
These figures agree well with those obtained from the YB and ASTFIT pipelines.

Concerning differences between pipelines we obtain for the grid-based results from BASTA and YB median absolute differences (relative to BASTA) of $4.9\%$ in mass, $1.7\%$ in radius, $9.7\%$ in age, and $0.7\%$ in density. Differences in physics between the grids used by YB and BASTA cause the YB mass or radius estimates to be larger than those of BASTA. Two of the three grids of models used by YB were constructed with diffusion, as does the BASTA grid for $M\lesssim1.15\,\rm  M_{\odot}$. For models of a given mass, those with diffusion tend to be of lower temperature and lower luminosity than those without diffusion. This means that in a given temperature range, models with diffusion will have a higher mass than models without diffusion, causing the type of difference that we see between YB and BASTA results because the masses of the stars analysed here are predominately higher than $1.15\, \rm M_{\odot}$. Since density is related to $\Delta\nu$, a higher mass will automatically result in a higher radius. The third grid of models did not include diffusion, but since the YB results were determined from a distribution function that had models from all three grids, the net result was somewhat higher masses.

The final set of recommended parameters for our target sample is taken from the BASTA pipeline; these can be found in \tref{tab:model_values} from IRFM inputs and in \tref{tab:model_values2} from SPC inputs.

\subsection{EPIC 204926239}
\label{sec:239}

This target was found by \citet[][]{1999AJ....117..354D} to have an $84\%$ probability of being a member of the Upper Sco association (USa). \citet[][]{2012ApJ...746..154P} estimated an age of around $11$ Myr for the association. Adopting the stellar parameters estimated by \citet[][]{2012ApJ...746..154P}, with a mass of $1.5\, \rm M_{\odot}$, the target would be contracting convectively as a T-Tauri star in its pre-main-sequence (PMS) phase \citep[][]{1998ApJ...507L.141M,2010aste.book.....A}. Such a star would not reach the classical instability strip where the Herbig Ae/Be stars reside, but should oscillate in a solar-like manner from its extensive convective envelope \citep[][]{2005JApA...26..171S}. While our detection of oscillations would be exciting if the star were in its PMS phase, we find that this is likely not the case. First, if we adopt the parameters estimated by \citet[][]{2012ApJ...746..154P} and assume the standard scaling relation for \numax extends to the PMS, a \numax-value of ${\sim}1283\, \rm\mu Hz$ is predicted --- we found $\numax =747\pm 31\, \rm\mu Hz$. Secondly, T-Tauri stars are generally found to be very active, which should render low-amplitude solar-like oscillations difficult to observe. We find, however, only indications of low-amplitude variability that we ascribe to surface activity.
All things considered, we find it unlikely that EPIC 204926239 should be a PMS solar-like oscillator. Model grids were therefore not extended to the PMS phase in modelling this target.


\subsection{Seismic distances}
\label{sec:dist}

With the seismic solution for the stellar radii and an angular diameter from the IRFM, we can estimate the distance to a given target as follows: 
\begin{equation}
D_{\rm seis} = C \frac{2R_{\rm seis}}{\theta_{\rm IRFM}}\, ,
\end{equation} 
where $C$ is the conversion factor to parsec\footnote{We adopt $\rm 1\,AU = 149.5978707\, \times 10^{6}\, km$} \citep[see][]{2012ApJ...757...99S,2014MNRAS.445.2758R}.

\begin{figure}
\centering
\includegraphics[width=\columnwidth]{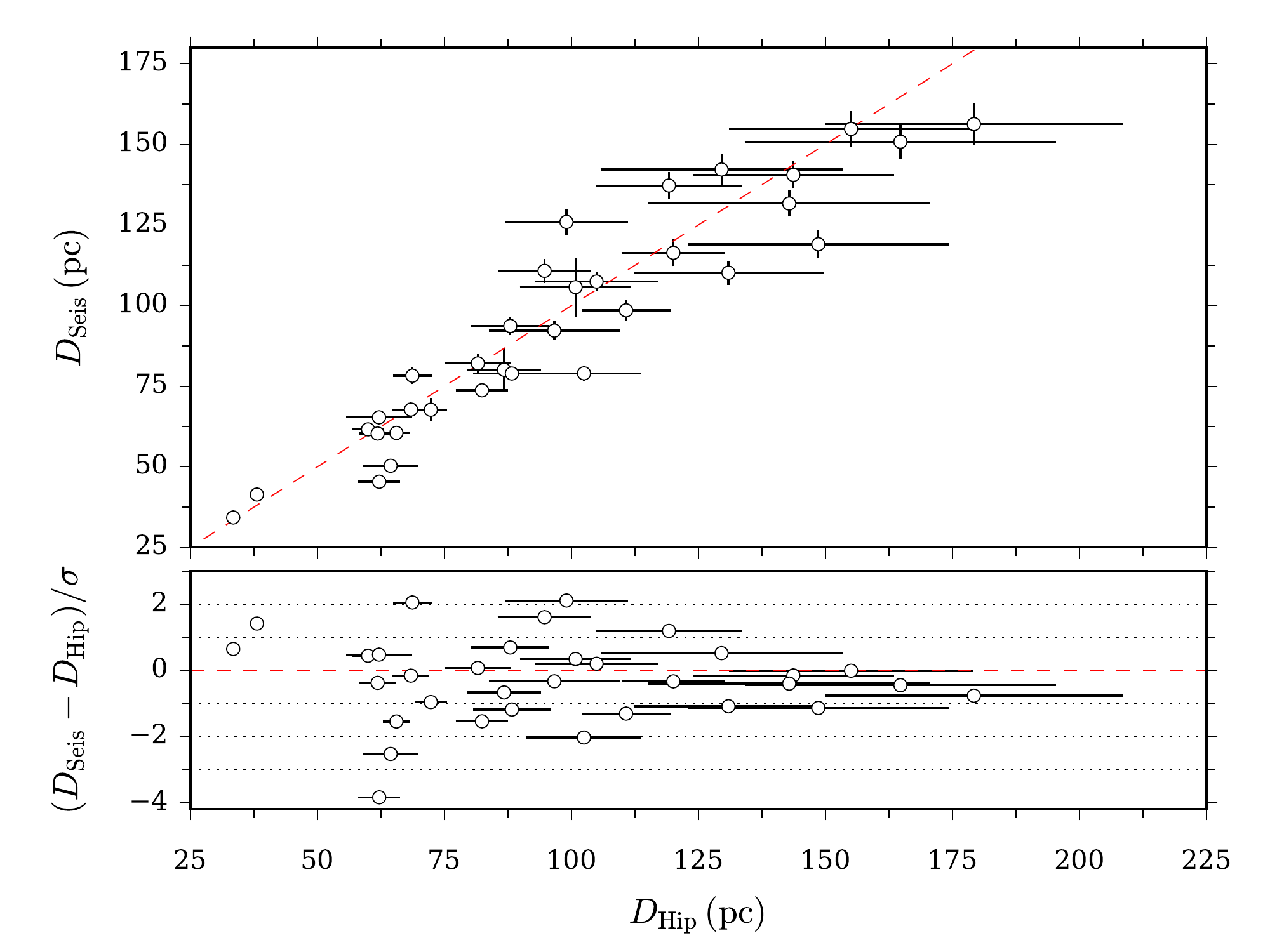}
\caption{Top: Comparison between distances from seismology ($D_{\rm seis}$) using grid-based results from \tref{tab:model_values} and from \textit{Hipparcos} parallaxes ($D_{\rm Hip}$). Bottom: Normalised distance difference (divided by the uncertainty on the difference) against \textit{Hipparcos} distances. We have omitted indicating the uncertainty on the normalised difference which by definition is 1 for all targets. The dashed lines show the $1:1$ distance relation; the dotted lines in the bottom panel indicate the $1\sigma$ increments in the normalised differences. }
\label{fig:distances}
\end{figure}

In \fref{fig:distances} we show the comparison between seismic and parallax distances. Overall we see a good agreement in the sense that no general systematic trend is seen in the differences with distance; from a Wilcoxon signed-rank test we find that the results are consistent with the null hypothesis of a symmetric distribution around zero in the differences \citep[][]{WilcoxonTest,Barlow:0471922951}. We find 14 targets with a difference beyond $\pm1\sigma$, of these 5 have differences beyond $\pm2\sigma$ --- these values are slightly larger than one would expect from a normal distribution of the residuals. The reduced $\chi_{\rm R}^2$ is ${\sim}1.7$ which could indicate that some uncertainties are underestimated, either for the parallax or seismic estimates. For the targets with modelling from individual frequencies distances are fully consistent with those from the grid-based modelling. 

With these three early campaigns from K2 we can already match the number of targets for which such a comparison was possible in the nominal \kp mission, that is, the 22 targets studied in \citet[][]{2012ApJ...757...99S} for which both \textit{Hipparcos} and seismic distances were available.

\begin{table*}
\renewcommand{\arraystretch}{1}
\centering 
\caption{Derived parameters from the seismic modelling with BASTA, with inputs from the IRFM, of the K2 targets with detected oscillations in SC from campaigns 1-3. For the incorporation of systematic uncertainties from different pipelines and input parameters see \sref{sec:comp} (see also \citealt[][]{2015MNRAS.452.2127S}). ``Cam.'' gives the K2 campaign; ``Source'' indicate whether the modelling used a grid-based approach (grid) or used the individual frequencies (indv.).  } 
\label{tab:model_values} 
\tabsize
\begin{tabular}{lccccccccccc} 
\toprule 
EPIC & Cam. & HIP. ID & Source & $\rm Mass$  & $\rm Radius$ & $\rm Density$ & $\log g$         & $\rm Age$ & $\rm Distance$ & \teff & $\rm [Fe/H]$\\ 
   & & & & $\rm (M_{\sun})$& $\rm (R_{\sun})$ & $(\rm g/cm^3)$& ($\rm cgs$; dex) & (Gyr)     & (pc)       & (K)           & (dex)  \\
\midrule
$201367296$                  & 1 & $58093$   & grid  &  $1.15 \pm 0.04$  & $1.73 \pm 0.04$ & $0.311 \pm 0.015$ & $4.019 \pm 0.014$ & $6.55  \pm 0.73$  & $61.60  \pm 1.94$ & $5731 \pm 65$   &  $0.21  \pm 0.08$ \\
$201367904$                  & 1 & $58191$   & grid  &  $1.30 \pm 0.05$  & $2.08 \pm 0.05$ & $0.204 \pm 0.010$ & $3.914 \pm 0.013$ & $3.75  \pm 0.42$  & $137.22 \pm 4.23$ & $6225 \pm 78$   &  $-0.02 \pm 0.08$ \\
$201820830$                  & 1 & $55778$   & grid  &  $1.21 \pm 0.09$  & $1.71 \pm 0.05$ & $0.339 \pm 0.017$ & $4.052 \pm 0.016$ & $4.49  \pm 1.17$  & $116.38 \pm 4.14$ & $6264 \pm 104$  &  $-0.02 \pm 0.11$ \\
$201860743$                  & 1 & $57676$   & grid  &  $1.17 \pm 0.06$  & $1.90 \pm 0.05$ & $0.240 \pm 0.012$ & $3.947 \pm 0.014$ & $5.59  \pm 0.98$  & $82.07  \pm 2.87$ & $6004 \pm 130$  &  $-0.02 \pm 0.08$ \\
$204506926$                  & 2 & $81413$   & grid  &  $1.03 \pm 0.04$  & $1.38 \pm 0.03$ & $0.545 \pm 0.025$ & $4.166 \pm 0.013$ & $11.11 \pm 1.73$  & $78.98  \pm 2.25$ & $5562 \pm 71$   &  $0.38  \pm 0.08$ \\
$\,\,\,\,\,\,\,\,\,\,\cdots$ & 2 & $81413$   & indv. &  $1.04 \pm 0.04$  & $1.39 \pm 0.02$ & $0.537 \pm 0.006$ & $4.164 \pm 0.007$ & $10.73 \pm 1.65$  & $79.49  \pm 1.74$ & $5575 \pm 65$   &  $0.38  \pm 0.11$ \\
$204356572$                  & 2 & $80374$   & grid  &  $1.19 \pm 0.09$  & $1.50 \pm 0.04$ & $0.496 \pm 0.025$ & $4.159 \pm 0.018$ & $3.88  \pm 1.64$  & $80.08  \pm 6.82$ & $6342 \pm 182$  &  $-0.07 \pm 0.11$ \\
$204550630$                  & 2 & $81235$   & grid  &  $1.18 \pm 0.08$  & $1.37 \pm 0.03$ & $0.637 \pm 0.032$ & $4.231 \pm 0.019$ & $3.62  \pm 1.67$  & $105.70 \pm 9.13$ & $6342 \pm 195$  &  $-0.02 \pm 0.08$ \\
$204624076$                  & 2 & $80756$   & grid  &  $1.21 \pm 0.05$  & $1.82 \pm 0.04$ & $0.280 \pm 0.012$ & $3.997 \pm 0.014$ & $4.23  \pm 0.56$  & $107.44 \pm 3.01$ & $6342 \pm 71$   &  $-0.13 \pm 0.11$ \\
$204926239$                  & 2 & $79606$   & grid  &  $1.45 \pm 0.07$  & $2.46 \pm 0.05$ & $0.137 \pm 0.006$ & $3.816 \pm 0.013$ & $2.71  \pm 0.41$  & $142.19 \pm 4.71$ & $6238 \pm 156$  &  $0.10  \pm 0.08$ \\
$205917956$                  & 3 & $111312$  & grid  &  $1.03 \pm 0.11$  & $2.54 \pm 0.12$ & $0.089 \pm 0.005$ & $3.642 \pm 0.012$ & $7.86  \pm 2.80$  & $67.67  \pm 3.59$ & $5146 \pm 65$   &  $-0.35 \pm 0.06$ \\
$205962429$                  & 3 & $110537$  & grid  &  $1.17 \pm 0.03$  & $1.80 \pm 0.03$ & $0.283 \pm 0.011$ & $3.995 \pm 0.011$ & $4.76  \pm 0.41$  & $140.54 \pm 4.31$ & $6160 \pm 52$   &  $-0.30 \pm 0.08$ \\
$\,\,\,\,\,\,\,\,\,\,\cdots$ & 3 & $110537$  & indv. &  $1.16 \pm 0.02$  & $1.80 \pm 0.02$ & $0.283 \pm 0.006$ & $3.994 \pm 0.006$ & $4.96  \pm 0.27$  & $140.54 \pm 3.90$ & $6251 \pm 58$   &  $-0.18 \pm 0.08$ \\
$205967173$                  & 3 & $109822$  & grid  &  $0.96 \pm 0.09$  & $2.66 \pm 0.12$ & $0.072 \pm 0.004$ & $3.574 \pm 0.014$ & $10.71 \pm 3.15$  & $41.38  \pm 2.20$ & $5029 \pm 52$   &  $-0.30 \pm 0.08$ \\
$205974115$                  & 3 & $110689$  & grid  &  $1.17 \pm 0.04$  & $1.37 \pm 0.02$ & $0.644 \pm 0.025$ & $4.233 \pm 0.012$ & $3.66  \pm 0.53$  & $78.92  \pm 1.96$ & $6186 \pm 65$   &  $-0.13 \pm 0.08$ \\
$205979004$                  & 3 & $110454$  & grid  &  $1.12 \pm 0.05$  & $2.24 \pm 0.06$ & $0.141 \pm 0.007$ & $3.787 \pm 0.013$ & $5.40  \pm 0.73$  & $119.01 \pm 4.38$ & $5822 \pm 123$  &  $-0.35 \pm 0.08$ \\
$205995584$                  & 3 & $110518$  & grid  &  $1.31 \pm 0.07$  & $1.72 \pm 0.04$ & $0.363 \pm 0.016$ & $4.083 \pm 0.013$ & $2.87  \pm 0.71$  & $60.50  \pm 1.87$ & $6433 \pm 65$   &  $-0.07 \pm 0.08$ \\
$206009487$                  & 3 & $111892$  & grid  &  $1.07 \pm 0.04$  & $1.51 \pm 0.03$ & $0.435 \pm 0.018$ & $4.105 \pm 0.012$ & $6.66  \pm 0.98$  & $65.33  \pm 1.92$ & $6017 \pm 71$   &  $-0.24 \pm 0.08$ \\
$\,\,\,\,\,\,\,\,\,\,\cdots$ & 3 & $111892$  & indv. &  $1.08 \pm 0.04$  & $1.52 \pm 0.02$ & $0.432 \pm 0.008$ & $4.105 \pm 0.008$ & $6.45  \pm 0.86$  & $65.85  \pm 1.49$ & $6030 \pm 71$   &  $-0.24 \pm 0.08$ \\
$206064678$                  & 3 & $109672$  & grid  &  $0.92 \pm 0.03$  & $0.95 \pm 0.02$ & $1.489 \pm 0.062$ & $4.441 \pm 0.013$ & $9.42  \pm 2.01$  & $50.32  \pm 1.22$ & $5315 \pm 71$   &  $0.26  \pm 0.06$ \\
$\,\,\,\,\,\,\,\,\,\,\cdots$ & 3 & $109672$  & indv. &  $0.95 \pm 0.02$  & $0.97 \pm 0.01$ & $1.474 \pm 0.011$ & $4.443 \pm 0.004$ & $7.30  \pm 1.25$  & $51.27  \pm 0.96$ & $5393 \pm 52$   &  $0.38  \pm 0.06$ \\
$206064711$                  & 3 & $108692$  & grid  &  $1.28 \pm 0.05$  & $1.57 \pm 0.03$ & $0.462 \pm 0.018$ & $4.149 \pm 0.011$ & $2.57  \pm 0.38$  & $67.74  \pm 1.90$ & $6602 \pm 65$   &  $-0.18 \pm 0.08$ \\
$206070413$                  & 3 & $111534$  & grid  &  $1.25 \pm 0.06$  & $1.53 \pm 0.04$ & $0.493 \pm 0.022$ & $4.165 \pm 0.013$ & $2.87  \pm 0.64$  & $110.18 \pm 3.73$ & $6511 \pm 71$   &  $-0.13 \pm 0.11$ \\
$206078331$                  & 3 & $108468$  & grid  &  $0.96 \pm 0.04$  & $1.13 \pm 0.03$ & $0.922 \pm 0.036$ & $4.308 \pm 0.011$ & $8.51  \pm 1.63$  & $34.28  \pm 1.05$ & $5848 \pm 65$   &  $-0.18 \pm 0.11$ \\
$\,\,\,\,\,\,\,\,\,\,\cdots$ & 3 & $108468$  & indv. &  $0.96 \pm 0.04$  & $1.13 \pm 0.02$ & $0.938 \pm 0.012$ & $4.314 \pm 0.008$ & $8.26  \pm 1.60$  & $34.10  \pm 0.83$ & $5861 \pm 71$   &  $-0.18 \pm 0.11$ \\
$206088888$                  & 3 & $111376$  & grid  &  $1.10 \pm 0.05$  & $1.31 \pm 0.03$ & $0.691 \pm 0.028$ & $4.245 \pm 0.011$ & $5.14  \pm 1.10$  & $92.24  \pm 2.98$ & $6017 \pm 65$   &  $-0.02 \pm 0.08$ \\
$\,\,\,\,\,\,\,\,\,\,\cdots$ & 3 & $111376$  & indv. &  $1.10 \pm 0.04$  & $1.31 \pm 0.02$ & $0.681 \pm 0.011$ & $4.240 \pm 0.007$ & $5.35  \pm 0.94$  & $92.67  \pm 2.41$ & $6017 \pm 58$   &  $-0.07 \pm 0.08$ \\
$206094605$                  & 3 & $110065$  & grid  &  $1.37 \pm 0.12$  & $2.00 \pm 0.07$ & $0.243 \pm 0.010$ & $3.976 \pm 0.014$ & $2.84  \pm 0.80$  & $156.26 \pm 6.58$ & $6511 \pm 97$   &  $0.04  \pm 0.11$ \\
$206107253$                  & 3 & $110217$  & grid  &  $1.18 \pm 0.05$  & $1.31 \pm 0.03$ & $0.739 \pm 0.027$ & $4.274 \pm 0.011$ & $2.78  \pm 0.70$  & $93.66  \pm 2.90$ & $6485 \pm 65$   &  $-0.13 \pm 0.11$ \\
$206108325$                  & 3 & $110902$  & grid  &  $1.27 \pm 0.05$  & $2.14 \pm 0.05$ & $0.184 \pm 0.009$ & $3.881 \pm 0.015$ & $3.69  \pm 0.44$  & $131.64 \pm 4.02$ & $6303 \pm 97$   &  $-0.18 \pm 0.11$ \\
$206184719$                  & 3 & $111843$  & grid  &  $1.45 \pm 0.06$  & $2.78 \pm 0.07$ & $0.094 \pm 0.005$ & $3.708 \pm 0.013$ & $2.63  \pm 0.26$  & $78.26  \pm 2.68$ & $5926 \pm 97$   &  $-0.02 \pm 0.11$ \\
$206245055$                  & 3 & $111332$  & grid  &  $0.91 \pm 0.04$  & $0.94 \pm 0.02$ & $1.531 \pm 0.060$ & $4.447 \pm 0.012$ & $5.94  \pm 1.68$  & $60.30  \pm 1.77$ & $5913 \pm 65$   &  $-0.41 \pm 0.08$ \\
$\,\,\,\,\,\,\,\,\,\,\cdots$ & 3 & $111332$  & indv. &  $0.90 \pm 0.04$  & $0.94 \pm 0.01$ & $1.499 \pm 0.016$ & $4.440 \pm 0.007$ & $6.48  \pm 1.65$  & $60.50  \pm 1.52$ & $5913 \pm 65$   &  $-0.41 \pm 0.11$ \\
$206289767$                  & 3 & $109899$  & grid  &  $1.34 \pm 0.06$  & $2.06 \pm 0.05$ & $0.217 \pm 0.009$ & $3.939 \pm 0.014$ & $3.27  \pm 0.49$  & $150.81 \pm 5.32$ & $6342 \pm 91$   &  $0.04  \pm 0.11$ \\
$206368174$                  & 3 & $110002$  & grid  &  $1.33 \pm 0.09$  & $1.79 \pm 0.05$ & $0.325 \pm 0.013$ & $4.052 \pm 0.012$ & $3.16  \pm 0.88$  & $110.75 \pm 3.79$ & $6251 \pm 65$   &  $0.10  \pm 0.06$ \\
$206371648$                  & 3 & $109951$  & grid  &  $0.83 \pm 0.02$  & $0.92 \pm 0.02$ & $1.488 \pm 0.084$ & $4.425 \pm 0.015$ & $14.05 \pm 0.99$  & $45.38  \pm 1.41$ & $5406 \pm 52$   &  $-0.13 \pm 0.08$ \\
$206445085$                  & 3 & $109836$  & grid  &  $1.19 \pm 0.05$  & $1.87 \pm 0.04$ & $0.256 \pm 0.011$ & $3.969 \pm 0.012$ & $6.21  \pm 0.77$  & $73.71  \pm 2.23$ & $5757 \pm 78$   &  $0.26  \pm 0.11$ \\
$\,\,\,\,\,\,\,\,\,\,\cdots$ & 3 & $109836$  & indv. &  $1.22 \pm 0.03$  & $1.88 \pm 0.02$ & $0.259 \pm 0.006$ & $3.975 \pm 0.006$ & $5.80  \pm 0.49$  & $74.06  \pm 1.80$ & $5783 \pm 78$   &  $0.32  \pm 0.11$ \\
$206453540$                  & 3 & $109783$  & grid  &  $1.40 \pm 0.08$  & $1.86 \pm 0.05$ & $0.303 \pm 0.012$ & $4.040 \pm 0.013$ & $2.39  \pm 0.61$  & $98.52  \pm 3.38$ & $6459 \pm 117$  &  $0.04  \pm 0.08$ \\
$206189649$                  & 3 & $110974$  & grid  &  $1.16 \pm 0.05$  & $1.99 \pm 0.05$ & $0.208 \pm 0.009$ & $3.905 \pm 0.012$ & $5.27  \pm 0.57$  & $125.92 \pm 4.13$ & $6043 \pm 71$   &  $-0.18 \pm 0.11$ \\
$206201061$                  & 3 & $110077$  & grid  &  $1.32 \pm 0.11$  & $1.93 \pm 0.06$ & $0.260 \pm 0.011$ & $3.991 \pm 0.013$ & $3.19  \pm 0.85$  & $154.76 \pm 5.61$ & $6446 \pm 110$  &  $-0.02 \pm 0.11$ \\
\bottomrule
\end{tabular} 
\begin{tablenotes}[normal]  
  \scriptsize 
	\item \textbf{NOTE}: We adopt the following solar parameters: $\rm M_{\sun} = 1.989\times 10^{33}\, \rm g$; $\rm R_{\sun} = 6.9599\times 10^{10}\, \rm cm$
 \end{tablenotes}  
\end{table*}  

\begin{table*}
\renewcommand{\arraystretch}{1}
\centering 
\caption{Derived parameters from the seismic modelling with BASTA, with inputs from the SPC, of the K2 targets with detected oscillations in SC from campaigns 1-3. For the incorporation of systematic uncertainties from different pipelines and input parameters see \sref{sec:comp} (see also \citealt[][]{2015MNRAS.452.2127S}). ``Cam.'' gives the K2 campaign; ``Source'' indicate whether the modelling used a grid-based approach (grid) or used the individual frequencies (indv.).  } 
\label{tab:model_values2} 
\tabsize
\begin{tabular}{lccccccccccc} 
\toprule 
EPIC & Cam. & HIP. ID & Source & $\rm Mass$  & $\rm Radius$ & $\rm Density$ & $\log g$         & $\rm Age$ & $\rm Distance$ & \teff & $\rm [Fe/H]$\\ 
   & & & & $\rm (M_{\sun})$& $\rm (R_{\sun})$ & $(\rm g/cm^3)$& ($\rm cgs$; dex) & (Gyr)     & (pc)       & (K)           & (dex)  \\
\midrule
$201367296$                  & 1 & $58093$   & grid  &  $1.15 \pm 0.04$  & $1.73 \pm 0.04$ & $0.311 \pm 0.015$ & $4.018 \pm 0.013$ & $6.63  \pm 0.77$  & $61.49  \pm 1.90$ & $5718 \pm 71$   &  $0.21  \pm 0.08$ \\
$201367904$                  & 1 & $58191$   & grid  &  $1.30 \pm 0.05$  & $2.08 \pm 0.05$ & $0.204 \pm 0.010$ & $3.916 \pm 0.013$ & $3.64  \pm 0.42$  & $137.42 \pm 4.24$ & $6251 \pm 78$   &  $-0.02 \pm 0.08$ \\
$201820830$                  & 1 & $55778$   & grid  &  $1.29 \pm 0.08$  & $1.74 \pm 0.05$ & $0.343 \pm 0.017$ & $4.063 \pm 0.016$ & $3.24  \pm 0.79$  & $118.00 \pm 4.24$ & $6420 \pm 71$   &  $-0.07 \pm 0.11$ \\
$201860743$                  & 1 & $57676$   & grid  &  $1.14 \pm 0.05$  & $1.88 \pm 0.05$ & $0.240 \pm 0.012$ & $3.945 \pm 0.014$ & $6.07  \pm 0.76$  & $81.55  \pm 2.81$ & $5926 \pm 78$   &  $-0.02 \pm 0.11$ \\
$204506926$                  & 2 & $81413$   & grid  &  $1.08 \pm 0.05$  & $1.41 \pm 0.03$ & $0.543 \pm 0.024$ & $4.172 \pm 0.013$ & $8.57  \pm 1.62$  & $80.52  \pm 2.40$ & $5705 \pm 78$   &  $0.38  \pm 0.08$ \\
$\,\,\,\,\,\,\,\,\,\,\cdots$ & 2 & $81413$   & indv. &  $1.09 \pm 0.05$  & $1.41 \pm 0.02$ & $0.541 \pm 0.006$ & $4.173 \pm 0.008$ & $8.32  \pm 1.53$  & $80.69  \pm 1.76$ & $5718 \pm 78$   &  $0.38  \pm 0.08$ \\
$204356572$                  & 2 & $80374$   & grid  &  $1.18 \pm 0.07$  & $1.50 \pm 0.04$ & $0.493 \pm 0.024$ & $4.157 \pm 0.015$ & $4.07  \pm 1.01$  & $80.08  \pm 6.77$ & $6316 \pm 78$   &  $-0.07 \pm 0.08$ \\
$204550630$                  & 2 & $81235$   & grid  &  $1.17 \pm 0.04$  & $1.37 \pm 0.03$ & $0.634 \pm 0.030$ & $4.228 \pm 0.014$ & $3.61  \pm 0.54$  & $105.70 \pm 8.95$ & $6199 \pm 65$   &  $-0.13 \pm 0.08$ \\
$204624076$                  & 2 & $80756$   & grid  &  $1.22 \pm 0.05$  & $1.83 \pm 0.04$ & $0.281 \pm 0.012$ & $4.000 \pm 0.014$ & $4.07  \pm 0.54$  & $107.62 \pm 3.01$ & $6381 \pm 71$   &  $-0.13 \pm 0.11$ \\
$204926239$                  & 2 & $79606$   & grid  &  $1.44 \pm 0.05$  & $2.45 \pm 0.05$ & $0.137 \pm 0.006$ & $3.817 \pm 0.012$ & $2.74  \pm 0.26$  & $141.84 \pm 4.65$ & $6225 \pm 71$   &  $0.10  \pm 0.08$ \\
$205917956$                  & 3 & $111312$  & grid  &  $1.06 \pm 0.11$  & $2.56 \pm 0.12$ & $0.089 \pm 0.005$ & $3.644 \pm 0.012$ & $7.04  \pm 2.63$  & $68.31  \pm 3.59$ & $5185 \pm 71$   &  $-0.35 \pm 0.08$ \\
$205962429$                  & 3 & $110537$  & grid  &  $1.15 \pm 0.04$  & $1.79 \pm 0.04$ & $0.281 \pm 0.011$ & $3.990 \pm 0.012$ & $5.25  \pm 0.63$  & $140.08 \pm 4.52$ & $6082 \pm 65$   &  $-0.24 \pm 0.08$ \\
$\,\,\,\,\,\,\,\,\,\,\cdots$ & 3 & $110537$  & indv. &  $1.17 \pm 0.02$  & $1.80 \pm 0.02$ & $0.281 \pm 0.005$ & $3.993 \pm 0.004$ & $5.03  \pm 0.18$  & $141.01 \pm 3.78$ & $6212 \pm 45$   &  $-0.13 \pm 0.06$ \\
$205967173$                  & 3 & $109822$  & grid  &  $0.95 \pm 0.08$  & $2.65 \pm 0.11$ & $0.073 \pm 0.004$ & $3.572 \pm 0.013$ & $11.14 \pm 3.00$  & $41.24  \pm 2.04$ & $5016 \pm 52$   &  $-0.30 \pm 0.08$ \\
$205974115$                  & 3 & $110689$  & grid  &  $1.15 \pm 0.04$  & $1.36 \pm 0.03$ & $0.641 \pm 0.025$ & $4.230 \pm 0.012$ & $4.10  \pm 0.76$  & $78.75  \pm 2.08$ & $6121 \pm 71$   &  $-0.07 \pm 0.08$ \\
$205979004$                  & 3 & $110454$  & grid  &  $1.07 \pm 0.06$  & $2.22 \pm 0.06$ & $0.139 \pm 0.007$ & $3.777 \pm 0.014$ & $6.45  \pm 0.99$  & $117.89 \pm 4.36$ & $5588 \pm 84$   &  $-0.35 \pm 0.08$ \\
$205995584$                  & 3 & $110518$  & grid  &  $1.31 \pm 0.07$  & $1.71 \pm 0.04$ & $0.363 \pm 0.016$ & $4.082 \pm 0.013$ & $2.89  \pm 0.72$  & $60.39  \pm 1.87$ & $6433 \pm 71$   &  $-0.07 \pm 0.08$ \\
$206009487$                  & 3 & $111892$  & grid  &  $1.03 \pm 0.05$  & $1.49 \pm 0.04$ & $0.436 \pm 0.018$ & $4.101 \pm 0.012$ & $7.97  \pm 1.25$  & $64.55  \pm 2.02$ & $5926 \pm 71$   &  $-0.24 \pm 0.08$ \\
$\,\,\,\,\,\,\,\,\,\,\cdots$ & 3 & $111892$  & indv. &  $1.04 \pm 0.05$  & $1.51 \pm 0.03$ & $0.429 \pm 0.007$ & $4.098 \pm 0.009$ & $7.58  \pm 1.21$  & $65.20  \pm 1.64$ & $5939 \pm 78$   &  $-0.24 \pm 0.08$ \\
$206064678$                  & 3 & $109672$  & grid  &  $1.01 \pm 0.04$  & $0.98 \pm 0.02$ & $1.483 \pm 0.060$ & $4.453 \pm 0.013$ & $4.23  \pm 1.65$  & $52.06  \pm 1.30$ & $5562 \pm 71$   &  $0.26  \pm 0.11$ \\
$\,\,\,\,\,\,\,\,\,\,\cdots$ & 3 & $109672$  & indv. &  $1.00 \pm 0.02$  & $0.98 \pm 0.01$ & $1.475 \pm 0.010$ & $4.451 \pm 0.004$ & $4.60  \pm 1.07$  & $52.06  \pm 0.97$ & $5562 \pm 65$   &  $0.21  \pm 0.08$ \\
$206064711$                  & 3 & $108692$  & grid  &  $1.27 \pm 0.05$  & $1.57 \pm 0.03$ & $0.460 \pm 0.017$ & $4.147 \pm 0.011$ & $2.74  \pm 0.45$  & $67.61  \pm 1.89$ & $6563 \pm 71$   &  $-0.18 \pm 0.08$ \\
$206070413$                  & 3 & $111534$  & grid  &  $1.23 \pm 0.07$  & $1.52 \pm 0.04$ & $0.490 \pm 0.022$ & $4.160 \pm 0.014$ & $3.22  \pm 0.78$  & $109.53 \pm 3.71$ & $6446 \pm 71$   &  $-0.13 \pm 0.08$ \\
$206078331$                  & 3 & $108468$  & grid  &  $0.94 \pm 0.04$  & $1.13 \pm 0.03$ & $0.924 \pm 0.036$ & $4.306 \pm 0.012$ & $9.26  \pm 1.82$  & $34.10  \pm 1.05$ & $5809 \pm 78$   &  $-0.18 \pm 0.11$ \\
$\,\,\,\,\,\,\,\,\,\,\cdots$ & 3 & $108468$  & indv. &  $0.95 \pm 0.04$  & $1.13 \pm 0.02$ & $0.936 \pm 0.012$ & $4.311 \pm 0.008$ & $8.91  \pm 1.82$  & $34.01  \pm 0.83$ & $5809 \pm 71$   &  $-0.18 \pm 0.11$ \\
$206088888$                  & 3 & $111376$  & grid  &  $1.08 \pm 0.06$  & $1.30 \pm 0.03$ & $0.692 \pm 0.028$ & $4.242 \pm 0.012$ & $5.94  \pm 1.36$  & $91.40  \pm 3.04$ & $5965 \pm 71$   &  $-0.02 \pm 0.08$ \\
$\,\,\,\,\,\,\,\,\,\,\cdots$ & 3 & $111376$  & indv. &  $1.09 \pm 0.05$  & $1.31 \pm 0.02$ & $0.679 \pm 0.011$ & $4.238 \pm 0.008$ & $5.76  \pm 1.04$  & $92.24  \pm 2.45$ & $5978 \pm 58$   &  $-0.02 \pm 0.08$ \\
$206094605$                  & 3 & $110065$  & grid  &  $1.40 \pm 0.11$  & $2.01 \pm 0.07$ & $0.243 \pm 0.011$ & $3.979 \pm 0.014$ & $2.65  \pm 0.70$  & $157.43 \pm 6.60$ & $6537 \pm 78$   &  $0.04  \pm 0.08$ \\
$206107253$                  & 3 & $110217$  & grid  &  $1.16 \pm 0.03$  & $1.31 \pm 0.02$ & $0.727 \pm 0.026$ & $4.267 \pm 0.010$ & $3.10  \pm 0.58$  & $93.66  \pm 2.63$ & $6290 \pm 65$   &  $-0.18 \pm 0.08$ \\
$206108325$                  & 3 & $110902$  & grid  &  $1.30 \pm 0.05$  & $2.15 \pm 0.05$ & $0.185 \pm 0.009$ & $3.887 \pm 0.014$ & $3.30  \pm 0.36$  & $132.20 \pm 4.17$ & $6433 \pm 71$   &  $-0.18 \pm 0.11$ \\
$206184719$                  & 3 & $111843$  & grid  &  $1.44 \pm 0.06$  & $2.78 \pm 0.08$ & $0.094 \pm 0.005$ & $3.706 \pm 0.013$ & $2.68  \pm 0.26$  & $78.18  \pm 2.71$ & $5822 \pm 84$   &  $-0.02 \pm 0.11$ \\
$206245055$                  & 3 & $111332$  & grid  &  $0.88 \pm 0.04$  & $0.93 \pm 0.02$ & $1.534 \pm 0.060$ & $4.444 \pm 0.012$ & $7.20  \pm 1.95$  & $59.73  \pm 1.76$ & $5835 \pm 71$   &  $-0.35 \pm 0.08$ \\
$\,\,\,\,\,\,\,\,\,\,\cdots$ & 3 & $111332$  & indv. &  $0.88 \pm 0.04$  & $0.94 \pm 0.01$ & $1.495 \pm 0.016$ & $4.436 \pm 0.007$ & $7.71  \pm 1.94$  & $60.11  \pm 1.52$ & $5848 \pm 71$   &  $-0.41 \pm 0.08$ \\
$206289767$                  & 3 & $109899$  & grid  &  $1.36 \pm 0.08$  & $2.06 \pm 0.06$ & $0.217 \pm 0.009$ & $3.940 \pm 0.013$ & $3.16  \pm 0.48$  & $151.25 \pm 5.49$ & $6368 \pm 78$   &  $0.04  \pm 0.11$ \\
$206368174$                  & 3 & $110002$  & grid  &  $1.32 \pm 0.08$  & $1.78 \pm 0.05$ & $0.325 \pm 0.013$ & $4.050 \pm 0.013$ & $3.35  \pm 0.97$  & $110.38 \pm 3.93$ & $6212 \pm 65$   &  $0.10  \pm 0.06$ \\
$206371648$                  & 3 & $109951$  & grid  &  $0.83 \pm 0.03$  & $0.93 \pm 0.02$ & $1.455 \pm 0.083$ & $4.421 \pm 0.016$ & $12.44 \pm 1.92$  & $45.82  \pm 1.47$ & $5549 \pm 65$   &  $-0.24 \pm 0.08$ \\
$206445085$                  & 3 & $109836$  & grid  &  $1.18 \pm 0.04$  & $1.86 \pm 0.04$ & $0.256 \pm 0.011$ & $3.968 \pm 0.011$ & $6.51  \pm 0.80$  & $73.47  \pm 2.19$ & $5718 \pm 71$   &  $0.26  \pm 0.11$ \\
$\,\,\,\,\,\,\,\,\,\,\cdots$ & 3 & $109836$  & indv. &  $1.22 \pm 0.03$  & $1.88 \pm 0.02$ & $0.259 \pm 0.005$ & $3.975 \pm 0.006$ & $5.94  \pm 0.46$  & $73.94  \pm 1.80$ & $5744 \pm 71$   &  $0.38  \pm 0.11$ \\
$206453540$                  & 3 & $109783$  & grid  &  $1.42 \pm 0.08$  & $1.87 \pm 0.05$ & $0.304 \pm 0.012$ & $4.044 \pm 0.011$ & $2.22  \pm 0.43$  & $98.83  \pm 3.33$ & $6524 \pm 78$   &  $-0.02 \pm 0.11$ \\
$206189649$                  & 3 & $110974$  & grid  &  $1.17 \pm 0.06$  & $2.00 \pm 0.05$ & $0.209 \pm 0.009$ & $3.907 \pm 0.012$ & $5.06  \pm 0.59$  & $126.30 \pm 3.98$ & $6082 \pm 78$   &  $-0.18 \pm 0.11$ \\
$206201061$                  & 3 & $110077$  & grid  &  $1.41 \pm 0.10$  & $1.97 \pm 0.06$ & $0.259 \pm 0.011$ & $3.997 \pm 0.013$ & $2.52  \pm 0.64$  & $158.37 \pm 5.85$ & $6589 \pm 71$   &  $-0.02 \pm 0.11$ \\
\bottomrule
\end{tabular} 
\begin{tablenotes}[normal]  
  \scriptsize 
	\item \textbf{NOTE}: We adopt the following solar parameters: $\rm M_{\sun} = 1.989\times 10^{33}\, \rm g$; $\rm R_{\sun} = 6.9599\times 10^{10}\, \rm cm$
 \end{tablenotes}  
\end{table*} 


\section{Future K2 campaigns}
\label{sec:fut}


\subsection{Noise and detectability}
\label{sec:noise}

To address the question of detectability of oscillations in future campaigns, it is essential to understand the noise characteristics of the observations and their relation to the a posteriori detectability. The targets observed in C1-3 were selected from the \textit{Hipparcos} catalogue under the criteria of having a relative parallax uncertainty below $15\%$ and a ${\geq}95\%$ probability of detecting solar-like oscillations above the long-cadence (LC) Nyquist-frequency of $\rm{\sim}283\, \mu Hz$. For these early campaigns we deliberately sampled an extended range of the cool part of the HR-diagram to better determine the detectability in different regimes. The prediction of detectability was made using the recipe of \citet[][]{2011ApJ...732...54C}. Based on the noise levels obtained for the C1 targets analysed by \citet[][]{2015PASP..127.1038C}, it was predicted that oscillations could be detected in $7$ targets, of which $6$ in the end showed indications of oscillations --- the $4$ detections that were solid are adopted in this work. As mentioned in \sref{sec:power} it was anticipated from the C1 detections that solar-like oscillations should be detected up to $\rm{\sim}2500\, \mu Hz$ from C3 onwards, assuming the shot noise in the time series would reduce by a factor of ${\sim}2.5$ (${\sim}6$ in power) from the C1 values. This assumption was made on the grounds of an increase in the pointing attitude frequency from C3 onwards. An additional complication in predicting detectability comes from the lack of a priori knowledge of the level of activity, which will impact the seismic amplitudes.

\begin{figure*}
    \centering
    \begin{subfigure}
        \centering
        \includegraphics[width=\columnwidth]{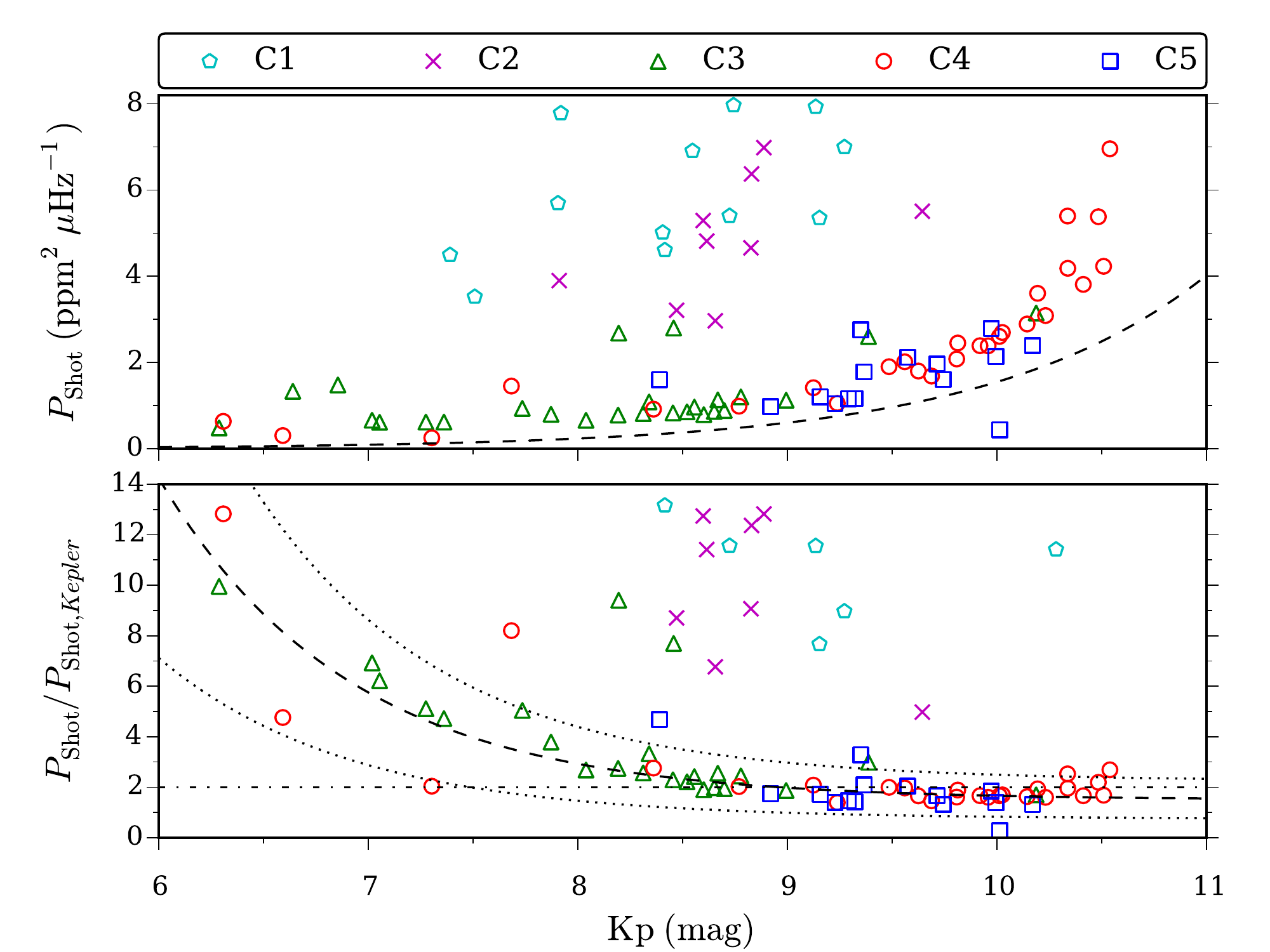}
    \end{subfigure}%
    \hfill 
    \begin{subfigure}
        \centering
        \includegraphics[width=\columnwidth]{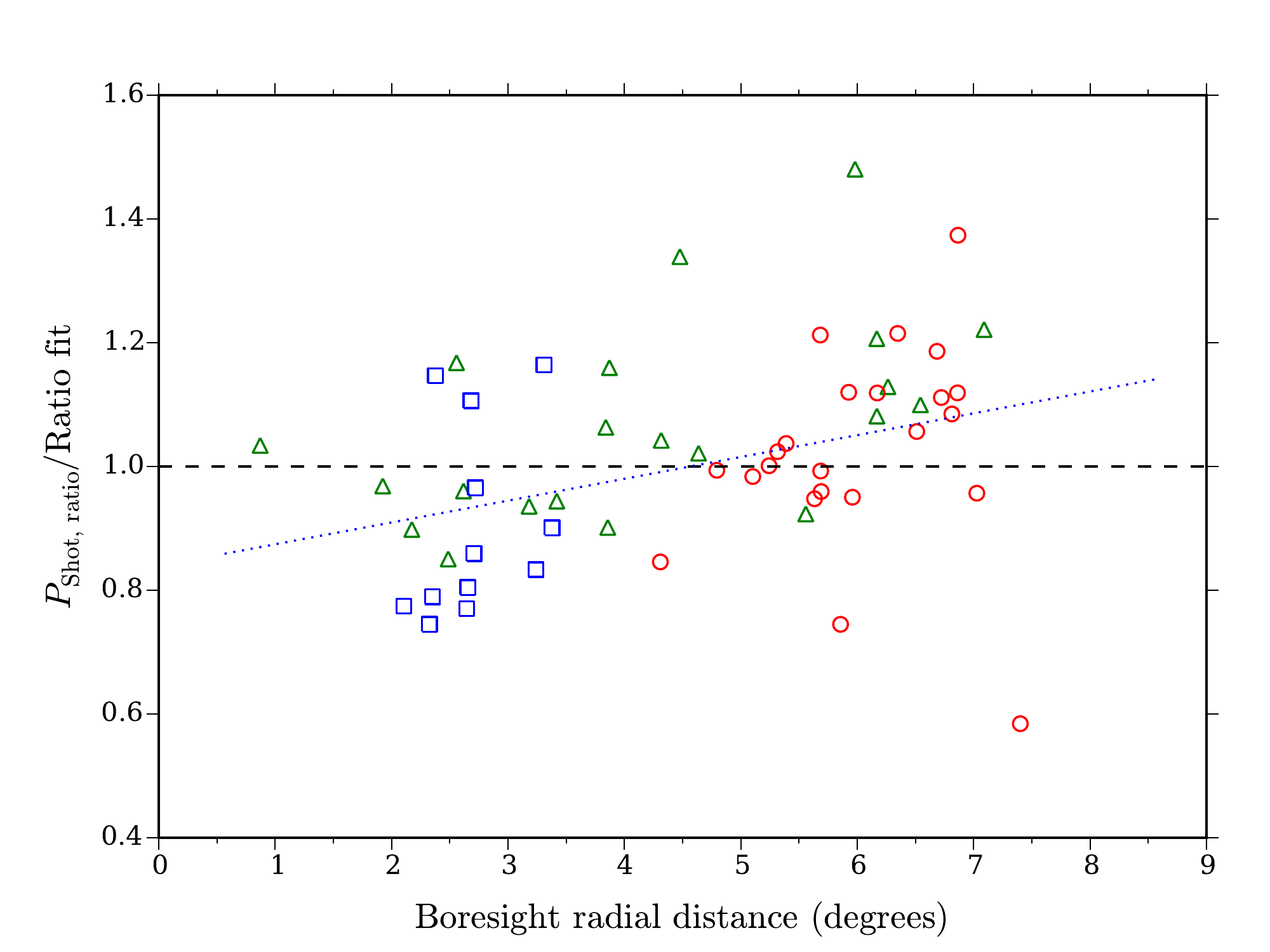}
    \end{subfigure}
    \caption{Left top: Shot-noise estimates for SC targets observed during K2 campaigns 1-5, measured as the robust mean power level above $8000\,\rm \mu Hz$. The dashed line gives the shot-noise floor from \citet[][]{2011ApJS..197....6G}. Left bottom: Shot noise divided by the \citet[][]{2011ApJS..197....6G} noise-floor relation; the dashed line shows the relation of this ratio against magnitude --- fitted ``by-eye'' to median binned ratios (not shown) of targets from campaigns 3-5. The dotted lines gives the dashed line times $0.5$ or $2$; we have also indicated the constant ratio of 2, which overall matches the observation well above $\Kp\sim 8$. Right: Shot-noise ratios divided by the ratio fit from the lower left panel against the radial boresight distance in degrees. The dotted line shows the linear relation which corresponds to the estimate Pearson's correlation between the plotted quantities.}
\label{fig:noise}
\end{figure*}

In the top left panel of \fref{fig:noise} we show the shot-noise estimates for SC targets observed during K2 campaigns 1-5. The targets from C4 and 5 are members of the open clusters M44, M67, and the Hyades. 
We estimated the shot noise from the median power density level above  $8000\,\rm \mu Hz$ divided by ${\sim}0.702$, which is the conversion factor between the median and mean level of a $\chi^2$ 2-d.o.f. noise distribution.
As seen, the noise levels significantly improved from C3 \citep[][]{2016van.cleve.pasp} and generally follow the trend in the noise floor (dashed line) from the nominal \kp mission by \citet[][]{2011ApJS..197....6G}.
The bottom left panel shows the shot-noise estimates divided by the \citet[][]{2011ApJS..197....6G} relation --- from $\rm 8 \lesssim Kp \lesssim 10.5$ the noise in K2 is only around a factor of 2 higher (in power) than in the nominal \kp mission. To get an estimate of the noise \kp-to-K2 noise ratio we have fitted ``by-eye'' a relation, given by the sum of two exponentially decaying functions and a constant offset, to the ratios shown in the lower left panel --- this approach is sufficient to enable a better prediction of the shot noise for future campaigns. Dividing the measured noise levels by the new noise relation we find a small positive correlation with boresight distance, meaning that targets observed near the \kp boresight are generally slightly less noisy than targets further away. This was also found in \citet[][]{2015ApJ...806...30L} and \citet[][]{2016van.cleve.pasp}, and can be seen as a natural consequence of the larger apparent movement of the targets on the CCD that lie far from the boresight.  

To compare the current detections against expectations we re-estimated the detectability of the C1-3 targets, largely using the detection recipe of \citet[][]{2011ApJ...732...54C}. We did this without the use of the spectroscopically or IRFM determined parameters in \tref{tab:parameter_values}, because such estimates typically only become available from follow-up observations after the fact. Hence, \teff were estimated from broad-band colours using the relations of \citet[][]{2010A&A...512A..54C}. Different from the \citet[][]{2011ApJ...732...54C} recipe we adopted the amplitude relation of \citet[][]{2011ApJ...743..143H}, and used as the solar bolometric \textsc{RMS} amplitude from \citet[][]{2009A&A...495..979M} of $2.53\pm 0.11$ ppm\footnote{the solar amplitude of $3.6$ ppm used in Huber et al. corresponds to the peak amplitude, which is $\sqrt{2}$ larger than the \textsc{RMS} value.}. As the amplitudes in \citet[][]{2011ApJ...743..143H} were estimated assuming a total visibility per order of $c=3.04$ this value is also used in predicting the detectability.
We used $\nu_{\rm max,\sun} = 3090\pm30\, \rm \mu Hz$, and $T_{\rm eff, \sun} = 5777$ K \citep[][]{2011ApJ...743..143H,2014ApJS..210....1C}, and the relation by \citet[][]{2011ApJ...743..143H} (from the SYD pipeline) to estimate \dnu from \numax.

\begin{figure*}
\centering
\includegraphics[width=2\columnwidth]{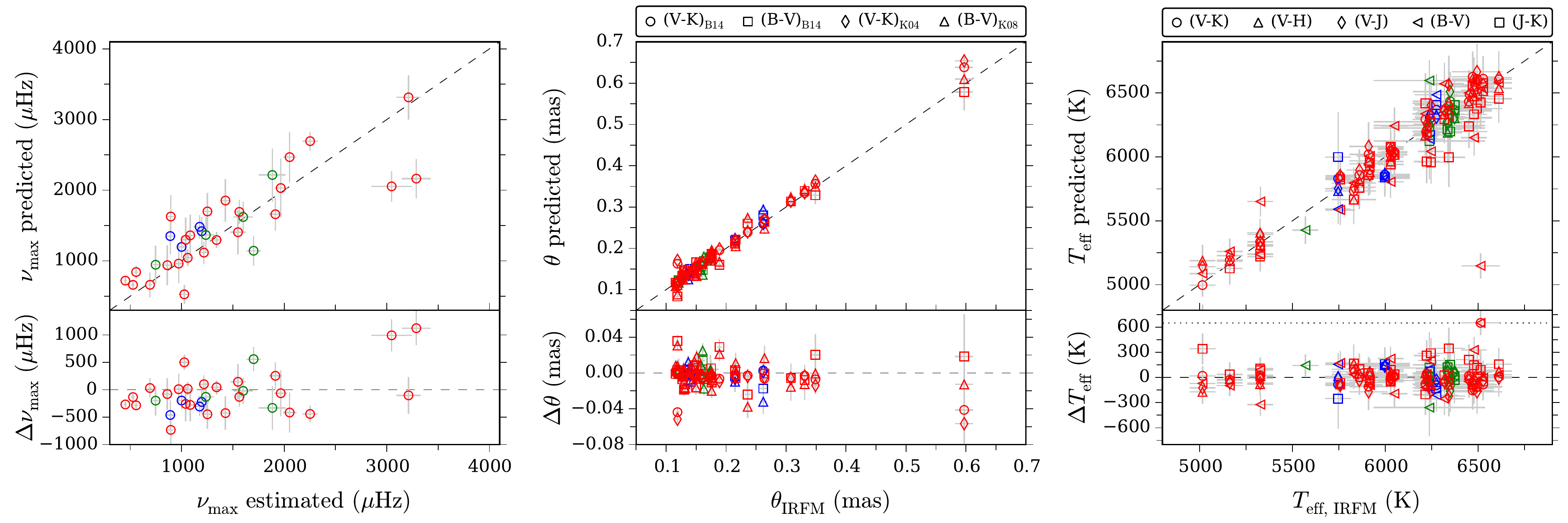}
\caption{Comparison between predicted and estimated values for parameters of interest in predicting detectability of seismic signals and the possibility of interferometric constraints. In each panel we have indicated the $1:1$ line for the compared parameters and indicated the campaign with the marker colour, where C1 is blue, C2 is green, and C3 is red. Left: Comparison between estimated and measured values of \numax for the sample of stars with positive detections of oscillations; see the text for the order of preference for the relation for the effective temperature \teff used in deriving \numax. Middle: Comparison between angular diameters $\theta$ estimated from broad-band colours using the relations of \citet[][]{2014AJ....147...47B} (B14), \citet[][]{2004A&A...426..297K} (K04), or \citet[][]{2008A&A...491..855K} (K08) and those derived from the IFRM using the seismic constraints on $\log g$. Right: Similar to the middle panel, but for the effective temperature \teff using different relations from \citet[][]{2010A&A...512A..54C}. If the difference exceeded 600 K we truncated the difference to this value, marked by the dotted line. For the two C2 targets with \teff uncertainties $>1000$ K we have truncated their errorbars to avoid cluttering the plot.}
\label{fig:predict_compare}
\end{figure*}
Depending on the availability of $JHK_S$ magnitudes from 2MASS we used (in order of preference) either the $(V-K_S)$, $(V-H)$, or $(V-J)$ relation. If none of these could be used we resorted to the $(B-V)$ relation. As an indicator of potential problems with the $V$-band magnitude used in the above \teff relations we computed, if possible, also \teff from the $(J-K)$ relation. Such a check is important, because the predicted value of \numax ultimately depends on the $V$-band magnitude via the luminosity estimate (similarly, predicted angular diameters would be affected).
$V$ and $B$ magnitudes were extracted from the Tycho2 catalogue \citep[][]{2000A&A...355L..27H}, and converted to the Johnson system using the relations in \citet[][]{2002AJ....124.1670M,2006AJ....131.2360M}, which are based on the work by \citet[][]{2000PASP..112..961B}. For all targets we assumed a metallicity of $\rm [Fe/H]=0.0\pm 0.1$ to propagate our general ignorance of this parameter into the \teff uncertainty, and for all magnitudes we assumed an uncertainty of $\pm0.02$ magnitudes. The various magnitudes were corrected for reddening based on the estimate of $E(B-V)$ at the (parallax) distance of a given target from the 3D dust map by \citet[][]{2015ApJ...810...25G}, and using extinction-to-reddening ratios ($R_X \equiv A_X / E(B-V)$) of $R_V=3.1\pm 0.1$ \citep[][]{1989ApJ...345..245C}, $R_K=0.355\pm0.1$, $R_J=0.88\pm0.1$, and $R_H=0.535\pm0.1$ \citep[][]{1999PASP..111...63F}.
Luminosities were derived from the \textit{Hipparcos} parallaxes by \citet[][]{2007A&A...474..653V}, with the bolometric correction from the relations by \citet[][]{1996ApJ...469..355F}, as presented in \citet[][]{2010AJ....140.1158T}. 
Masses were approximated from a simple mass-luminosity relation, specifically $L\propto M^{4\pm 0.5}$ \citep[see, \eg,][]{salaris2005evolution,2007MNRAS.382.1073M,2015AJ....149..131E}. 

In \fref{fig:predict_compare} we show the comparison of predicted and measured values of \numax, angular diameters (see \sref{sec:inter}), and \teff for our sample. In general we see a good agreement between predicted and measured values. In the \teff comparison one target can be identified where the $(J-K)$ estimate agrees with the IRFM, but those from $(V-K_S)$ and $(B-V)$ are off by more than 1000 K. This likely indicates a problem with the $V$-band magnitude and emphasises the importance of the $(J-K)$ sanity check of the temperature; indeed the \numax for this target is underestimated (${\sim}500\, \rm\mu Hz$ vs. a measured value of ${\sim}1000\, \rm\mu Hz$) from the affected \teff and luminosity. Two of the high \numax targets ($>3000\, \rm \mu Hz$), HIP 109672 and 109951, appear to be off in the predicted \numax. For these targets the \teff and $\theta$ from different relations agree with each other and with the results from the IRFM, but both display a mismatch between seismic and parallax distances (\fref{fig:distances}) --- this leads us to conclude that the parallax is off, and that this via $L$ affects \numax. 
We note that HIP 109951 is listed as a double star, which might have affected the parallax determination.

\begin{figure}
\centering
\includegraphics[width=\columnwidth]{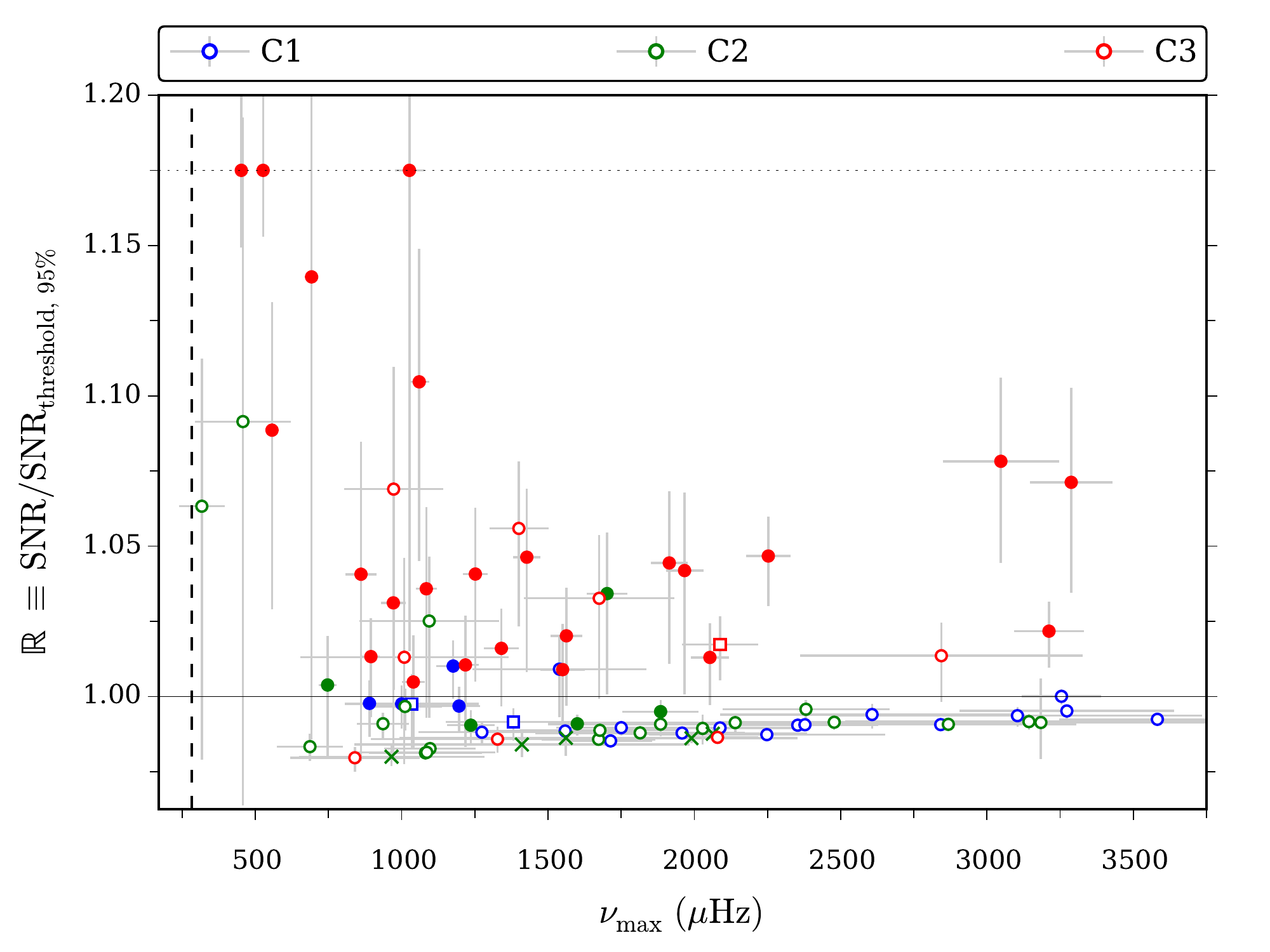}
\caption{Predicted detectability against \numax. The detectability is given by $\mathbb{R}$ as the estimated SNR over the threshold for a positive detection at the ${\geq}95\%$ level, \ie, a value above 1 indicates a predicted positive detection. The colours indicate the campaign of the target (see legend); filled markers indicate that a positive detection was made, vice versa for empty markers. The blue and red square markers indicate the marginal detections made in C1 \citep[][]{2015PASP..127.1038C} and C3; green crosses give the targets in C2 showing clear indications of Classical pulsations. Targets with a value of $\mathbb{R}>1.175$ have been truncated to this value, indicated by the dotted line. }
\label{fig:detect}
\end{figure}

In \fref{fig:detect} we show the predicted detectabilities for the C1-3 targets. The detectability is represented by the ratio $\mathbb{R}$ of the predicted signal-to-noise ratio (SNR) to the threshold SNR for a ${\geq}95\%$ probability detection, using the measured noise levels from \fref{fig:noise}; for targets with a measured \numax we have used that on the abscissa but the predicted value for calculating the detectability. We see that for C3 all targets with detected oscillations are indeed predicted to show oscillations. Three targets in C3 are predicted to yield no detections based on their measured noise level. For two of these we find bright close-by targets in the downloaded pixels that contaminate the light curves of the primary targets; the third is so bright that its flux spills out of the assigned pixels. We have thus detected oscillations in 24 out of 30 targets where we should hope to make a detection based on the shot-noise levels --- this corresponds to a success rate of $80\%$. For the remaining 6 targets in C3 we predict detectable oscillations, but find none. Most of these targets are found to be rather active, with power leaking into the power spectrum from low frequencies --- this will likely wash out the seismic signal, but not necessarily affect the shot-noise levels which were measured from frequencies above $8000\, \rm\mu Hz$. In addition, activity is known to attenuate the seismic signal, which further decreases the likelihood of making a detection \citep[][]{2011ApJ...732L...5C}. Activity is similarly found to be the culprit in the C2 non-detections where positive detections are predicted. 
All things considered we are confident that we understand the detectability of solar-like oscillations in K2.

\subsection{Future targets}
\label{sec:futtarg}

In the following we estimate the number of stars that will be observable in future K2 campaigns and have detectable oscillations in SC data, all of which were drawn from the \textit{Hipparcos} catalogue.
To ascertain which targets are on the detector we used the \texttt{K2fov} tool \citep[][]{fergal_mullally_2016_44283} as hosted on the website\footnote{\url{http://kasoc.phys.au.dk}} of the Kepler Asteroseismic Science Operation Center (KASOC). We note that the field pointings from C14 onwards are only approximate. For C6-8 we adopted the targets already selected for observations and for C10 we adopted the targets that were proposed (see \tref{tab:fut_cam}).

Fundamental parameters and predicted detectabilities were estimated as outlined \sref{sec:noise}. We assumed a duration of 80 days for all campaigns and required a detection probability ${\geq}95\%$ for a positive detection. We estimated the shot-noise level from the relation found in \sref{sec:noise}, but adopted a lower relative K2-to-\kp noise ratio of 2. Because the K2 Ecliptic Plane Input Catalog \citep[EPIC;][]{2015arXiv151202643H} is somewhat incomplete for later campaigns we computed \kp magnitudes (\Kp) following the recipe of \citet[][]{2011AJ....142..112B}. Here we adopted the transformation from Tycho $B_T$ and $V_T$ to Sloan $g$ and $r$, which were corrected for reddening using $R_g=3.3\pm 0.03$ and $R_r=2.31\pm 0.03$ \citep[][]{2013MNRAS.430.2188Y}.

\begin{figure}
\centering
\includegraphics[scale=0.45]{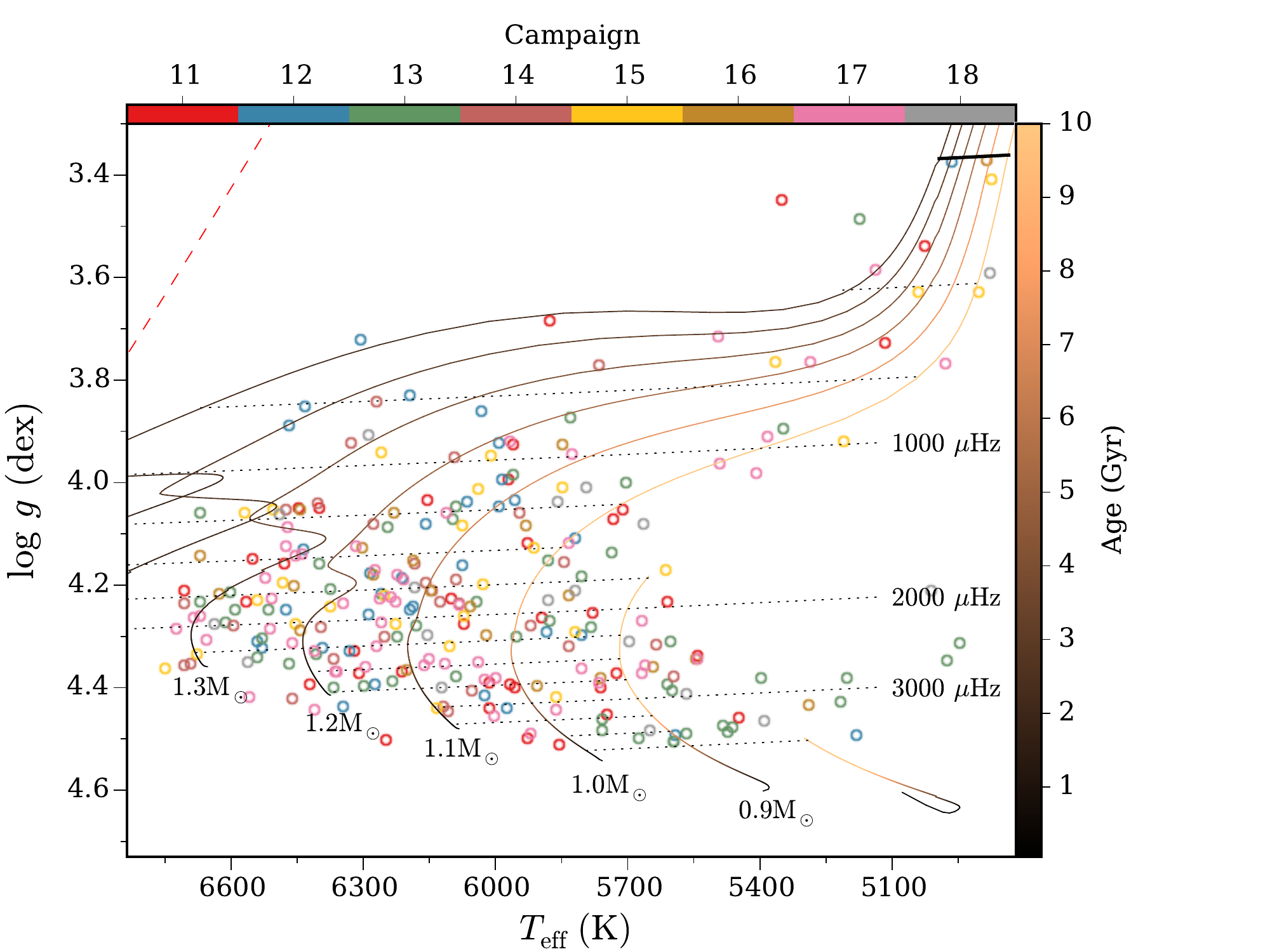}
\caption{Kiel-diagram of stars from K2 campaigns $11-18$ for which a detection of seismic power is predicted with a probability ${\geq}95\%$, and which have fractional parallax-uncertainties below $15\%$. Stellar evolutionary tracks have been calculated using GARSTEC adopting $\rm [Fe/H]=0$, with the model age indicated via the colour along the track. The top colour bar give the campaign of the targets. Indicated are lines of constant \numax, in increments of $\rm 250\, \mu Hz$, and we have specifically highlighted the iso-\numax lines at 1000, 2000, and 3000 $\rm \mu Hz$; the full black line gives the limit \numax at the LC nyquist frequency. The red dashed line gives the red-edge of the classical instability strip from \citet[][]{2000ASPC..210..215P}.}
\label{fig:HRdia_fut}
\end{figure}

The targets that are predicted to show detectable solar-like oscillations are show in a Kiel-diagram in \fref{fig:HRdia_fut} and listed in \tref{tab:fut_cam}.
We have trimmed the sample by requiring that the predicted \numax should be above the LC Nyquist frequency, and that the fractional uncertainty on the parallax is below $15\%$. The total number of targets from C6-18 approaches $431$. If we assume a success rate of $80\%$ then these, together with the targets analysed in the current work, will nearly match in numbers the $500$ main-sequence and sub-giant solar-like oscillators known to date from \kp \citep[][]{2011Sci...332..213C,2014ApJS..210....1C}.

The yields in C4 and 5 are relatively low because these campaigns were devoted to the study of the Hyades, Pleiades, M44, and M67 open clusters. From \fref{fig:noise} it is evident that these targets are, by and large, fainter than those observed in C3, which decreases the detectability. Moreover, many targets in the young cluster are fast rotators which also challenges the detection of oscillations.

\begin{figure*}
\centering
\includegraphics[scale=0.45]{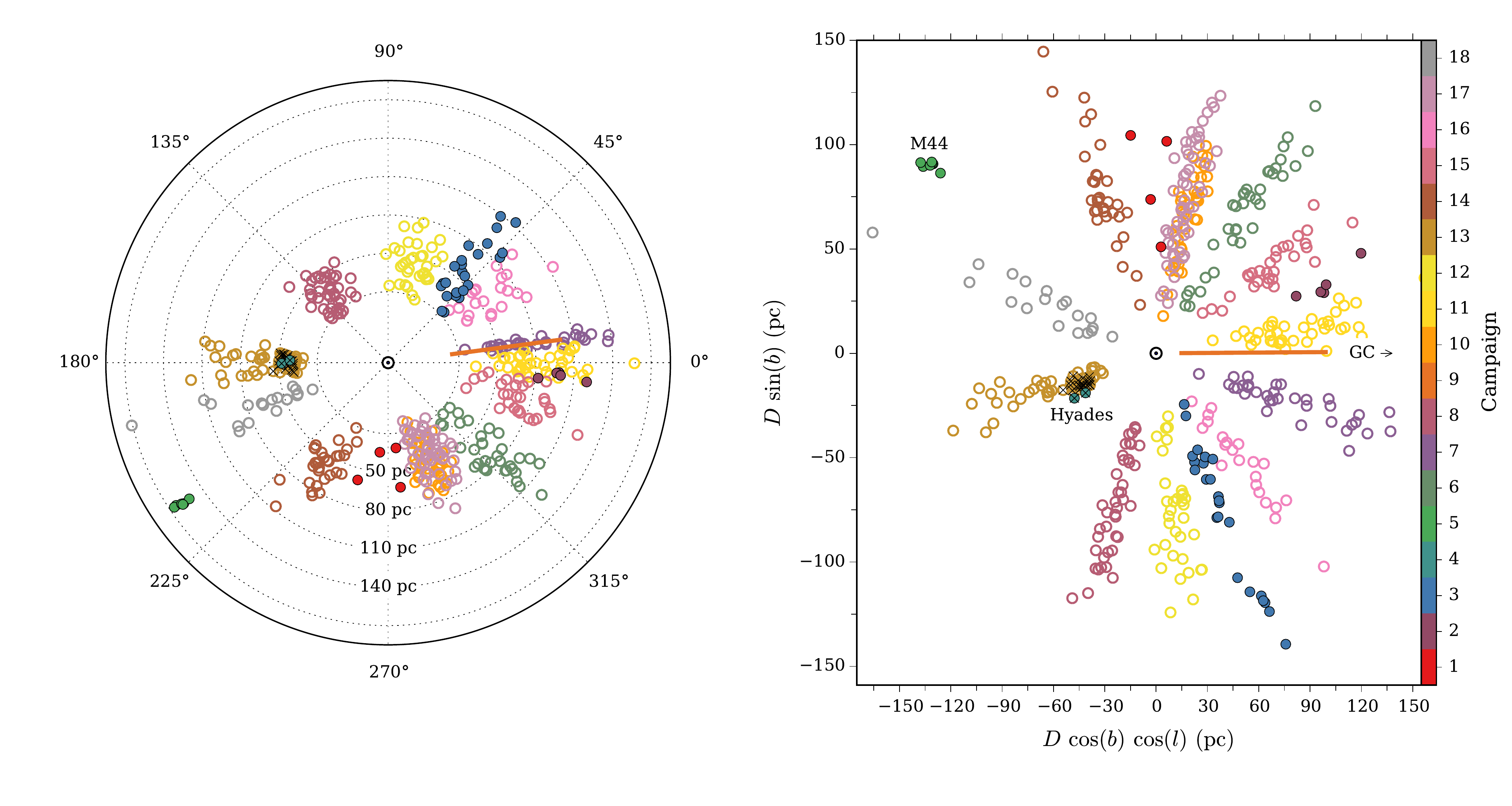}
\vspace{-15pt}
\caption{Sky positions in galactic coordinates of observed targets with positive seismic detections above the LC Nyquist (filled circular markers), future selected and proposed targets (empty circular markers). The targets from C4 and 13 belonging to the Hyades open cluster are further indicated with black crosses. For C10, we assumed that the 38 proposed targets will be selected. The future proposed targets comprise those for which we predict a positive seismic detections above the LC Nyquist. The M44 targets lack parallax distances, so these were drawn from a normal distribution as $\mathcal{N}(182, 5)$ pc \citep[][]{2009A&A...497..209V}; the targets identified in M67 fall outside the plot at a distance of ${\sim}832$ pc \citep[][]{2004MNRAS.347..101S} and in the approximate direction of M44. Left: Positions of targets in galactic longitude ($l$), latitude ($b$), and distance ($D$) from the Sun. The different colours indicate the K2 campaign (see colour bar in right panel). For C9, where no targets were proposed, we have indicated the direction with the coloured line. The galactic centre (GC) is in the direction of $l=0^{\circ}$. Right: Positions projected in the abscissa onto the $l=180^{\circ} \rightarrow 0^{\circ}$ line, with the direction of the GC to the right. The colour bar indicates the colour adopted for a given K2 campaign.}
\label{fig:sky_coor_gc}
\end{figure*}

In \fref{fig:sky_coor_gc} we show the sky positions of both current and future proposed targets, in galactic coordinates. As seen, K2 allows for an analysis using asteroseismology of the close solar neighbourhood. If all potential targets are eventually observed we may study the chemical evolution of the solar neighbourhood and place constraints on the age-metallicity relation of nearby field stars. 
A joint analysis of such a cohort would further allow us to thoroughly test seismic scaling relations, which are key components in, for instance, ensemble studies in galactic archaeology and population studies. Even better calibrations will be possible with coming Gaia parallaxes \citep[][]{2001A&A...369..339P}.

For standard aperture photometry and reduction using, \eg, the \pipe pipeline \citep[][]{2015ApJ...806...30L} it is only worth considering targets dimmer than $\Kp\gtrsim6.3$ mag because of the high level of saturation from brighter targets. For the brightest targets one may use a method as proposed in White et al. (in prep.) wherein photometry is performed from weighted sums of a relative small halo of unsaturated pixels around the saturated core of the target. This is indeed the method opted for in GO proposals 6081, 7081, and 8081 which focuses exactly on targets with $\Kp\lesssim6.3$, but in these proposals only for giants observed in LC. Alternatively, one may as outlined in \citet[][]{2016MNRAS.455L..36P} observe such bright stars indirectly from collateral smear photometry.
See \tref{tab:fut_cam} for the detection yields of current and future K2 campaigns.


\subsection{Interferometry}
\label{sec:inter}

It is interesting to look at the potential number of targets for which interferometry will be possible, because these targets will provide a near model-independent estimate of the stellar radius when combined with \textit{Hipparcos} or Gaia parallaxes. For this we assumed observations from the Precision Astronomical Visible Observations (PAVO) beam combiner \citep[][]{2008SPIE.7013E..24I} at the Center for High Angular Resolution Astronomy (CHARA) Array on Mt. Wilson Observatory, California  \citep[][]{2005ApJ...628..453T}.
The observing restrictions of PAVO can largely be summarised as a declination ${\gtrsim}-20^{\circ}$, an angular diameter ${\gtrsim}0.3$ mas, and a Johnson $R_J$-band magnitude ${\lesssim}8$.

For the \textit{Hipparcos} targets on the detector in C13-18 (for C14-18 from the currently proposed pointings) we estimated their angular diameters from the $(V-K)$ relation of \citet[][]{2014AJ....147...47B}, where we adopted the 2MASS $K_s$ for the $K$-band magnitude. If 2MASS photometry was absent we applied instead the $(B-V)$ relation by \citet[][]{2014AJ....147...47B}. As a cross-check we also derived the angular diameters from the $(V-K)$ and non-linear $(B-V)$ relations by \citet[][]{2004A&A...426..297K} and \citet[][]{2008A&A...491..855K}, see also \citet[][]{2012ApJ...760...32H}. 
\fref{fig:predict_compare} shows the overall excellent agreement between the estimates from these relations.
\begin{figure}
\centering
\includegraphics[scale=0.43]{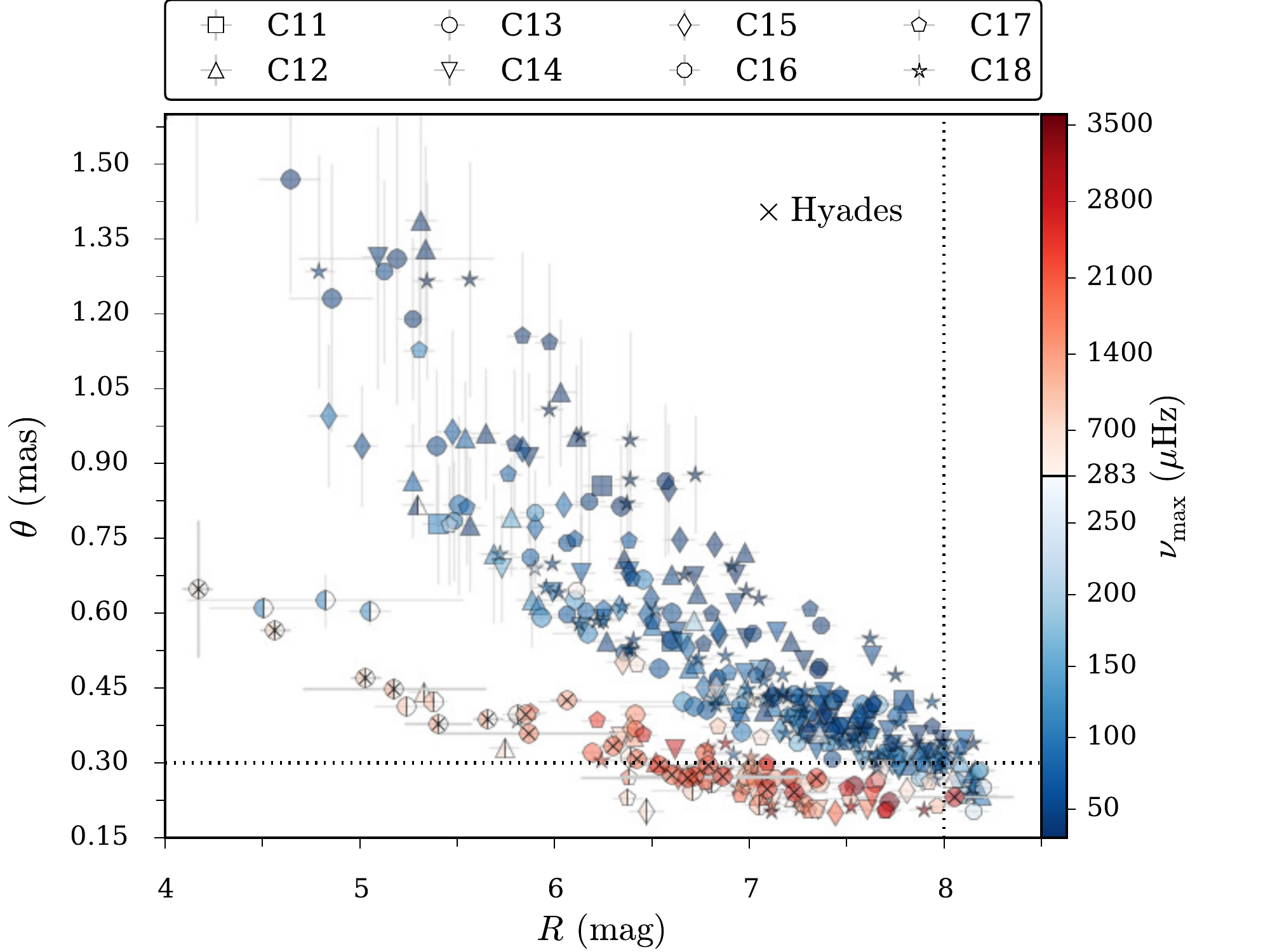}
\caption{Estimated angular diameters against $R$-band magnitudes for targets in C13-18 for which we anticipate detections of seismic excess power. All targets have declinations above $-20^{\circ}$, and no restrictions were put on parallax precision. The dotted lines indicate the approximate limits for sensible interferometric constraints from PAVO, hence targets fulfilling both constraints are found in the upper left quadrant. The colours indicate the predicted \numax value, where targets with \numax below (above) the LC Nyquist are given in blue (red). Red targets should thus be observed in SC. Note that the \numax range covered by the two partitions differ by more than an order of magnitude. For targets that are only half filled we predict a ratio $0.95<\mathbb{R}<1$. Targets that according to \citet[][]{1998A&A...331...81P} belongs to the Hyades open cluster are marked by crosses. }
\label{fig:interfer}
\end{figure}
To estimate the $R_J$ magnitudes we first used the relations by \citet[][]{2011AJ....142..112B} to convert Tycho $B_T$ and $V_T$ to analogues of Sloan $g$ and $r$ --- we then used the $R_J(g,r)$ relation by Lupton (2005)\footnote{\url{http://classic.sdss.org/dr5/algorithms/sdssUBVRITransform.html}}.
The $R_J$ band estimate was finally de-reddened using $R_R=2.32\pm0.1$ \citep[][]{1999PASP..111...63F} and $E(B-V)$ from \citet[][]{2015ApJ...810...25G} as determined in \sref{sec:noise}.

In \fref{fig:interfer} we show the targets from C13-18 with a declination above $-20^{\circ}$ that should show detectable oscillations, and with potential for interferometric inference. Targets rendered in blue are predicted to have \numax below the LC Nyquist frequency and vice versa for targets in red, which with $\numax>283\, \rm\mu Hz$ should be observed in SC mode. We find of the order ${\sim}30-40$ SC targets that will be suited for both asteroseismic and interferometric analysis. From \citet[][]{1998A&A...331...81P} several of these belong to the Hyades open cluster. An asteroseismic ensemble study of these, combined with independent constraints on radius from interferometry, would allow for tighter constraints to be put on the cluster distance and age. For an analysis of Hyades targets observed by K2 in C4 we refer to Lund et al. (2016, in press). 
Moreover, for the brightest targets in the sample we will have the possibility of conducting contemporaneous observations with the telescope of the Stellar Observations Network Group \citep[SONG;][]{2009ASPC..416..579G,2014IAUS..301...69G}.

\begin{table} 
\renewcommand{\arraystretch}{1}
\centering 
\caption{Overview of the number of targets observed or proposed in different campaigns together with the number of detections made. The parentheses in the ``targets'' column indicate that the values are our projected estimate of the number of targets that should show detectable oscillations. The values in parentheses in the ``detections'' indicate the projected yield, where we have assumed an $80\%$ success rate. On the bottom-line the values in parentheses include both targets from observed campaigns and those that are either selected or in the future.} 
\label{tab:fut_cam}
\mysize
\begin{tabular}{@{}lcccccc}  
\cmidrule[1.0pt](lr){2-7}
& Cam. & $\#$ targets & $\#$ detections & Proposal$^{\dagger}$ & PI$^{\dagger\dagger}$ & Notes\\ 
\cmidrule(lr){2-7}
\parbox[t]{1mm}{\multirow{8}{*}{\rotatebox[origin=c]{90}{\underline{Selected}}}}
& 1 & 23 & 4  & 1038 & Chaplin   &High noise\\
& 2 & 33 & 5  & 2038 & Chaplin   &High noise\\
& 3 & 33 & 24  & 3038 & Chaplin   &\\
& 4 & 31 & 2  & 4074 & Basu  & Hyades/Pleiades\\
& 5 & 51 & 12  & 5074 & Basu  & Praesepe/M67\\
& 6 & 35 & (28)  & 6039 & Davies  & North Galactic cap\\
& 7 & 34 & (27)  & 7039 & Davies  &Near galactic centre\\
& 8 & 41 & (33)  & 8002 & Campante  &\\
\cmidrule(lr){2-7}
\parbox[t]{1mm}{\multirow{5}{*}{\rotatebox[origin=c]{90}{\underline{Proposed}}}}
& 9 & 0 & 0 & --- &  --- &  Galactic centre\\
& 10 & (38) & (30)  & & Campante &  North Galactic cap\\
& 11 & (38) & (30)  & & Lund &  Galactic centre\\
& 12 & (34) & (27)  & & Lund &  South Galactic cap\\
& 13 & (53) & (42)  & & Lund &  Hyades\\
\cmidrule(lr){2-7}
\parbox[t]{1mm}{\multirow{5}{*}{\rotatebox[origin=c]{90}{\underline{Future}}}}
& 14 & (32) & (25)  & &  &  \\
& 15 & (29) & (23)   & &  &  \\
& 16 & (22) & (17)  & &  &  \\
& 17 & (56) & (44)  & &  &  North Galactic cap\\
& 18 & (19) & (15)  & &  &  \\
\cmidrule[0.5pt](lr){2-7}
&  & 281 (602) & 47 (388)  &  &  \\
\cmidrule[1.0pt](lr){2-7}
\end{tabular} 
\begin{tablenotes}[normal]  
     \scriptsize 
     \item $^{\dagger}$Proposal ID within the K2 Guest Observer (GO) program; $^{\dagger\dagger}$ Principal Investigator
 \end{tablenotes}  
\end{table} 

\section{Conclusion}
\label{sec:dis}

We have presented an asteroseismic analysis of 33 solar-like oscillators from SC observations during K2 campaigns 1-3. We find that the quality of the data from C3 onwards is sufficient for extraction of seismic parameters, and in addition to the average parameters \dnu and \numax one can peak-bag the frequency power spectra to recover individual frequencies. In terms of noise we find this to be better than what could have been hoped for, allowing us to detect oscillations beyond the solar \numax.    

Modelling was performed using different pipelines. The agreement between the values returned from these was excellent and only in a few cases beyond the uncertainties from the individual pipelines. For the grid-based pipelines these individual uncertainties were not significantly different from what could be obtained in the nominal \kp mission, because the uncertainty on \numax and \dnu is relatively insensitive to the duration of the observations. For individual frequencies the precision naturally improves with observing length, but for modes that are well resolved compared to the mode line width one may still obtain individual frequencies that are sufficiently precise for modelling.      

Concerning the comparison between seismic and parallax distances we see from \fref{fig:distances} that the dominant uncertainty is on the \textit{Hipparcos} distances. If one assumes an uncertainty on the parallax of $\rm {\sim}7\, \mu as$\footnote{\url{http://www.cosmos.esa.int/web/gaia/science-performance}}, which will likely be achieved from the Gaia mission \citep[][]{2015A&A...574A.115M} for the parameter range of the sample, such a comparison would provide a strong test of the results from seismic modelling and a better assessment of potential systematic differences. With an assumed parallax uncertainty of $\rm {\sim}7\, \mu as$, the dominant uncertainty in the comparison will shift to the other quantities needed to derive seismic distances, \ie, the seismic radii we wish to test and/or the uncertainties intrinsic to bolometric correction and angular diameter determinations. 

With K2 the possibility for obtaining independent radii measurements from interferometry is significantly improved, because the targets under study for solar-like oscillations are typically brighter than in the nominal \kp mission. In \sref{sec:inter} we found that several targets in C13-18, including members of the Hyades open cluster, will be apt for interferometric analysis. Because of the brightness of these targets we may further conduct contemporaneous observations with SONG --- a combined asteroseismic analysis of both the photometric light curve from K2 and the RV data from SONG would allow for a very detailed characterisation of a given star. 

Comparing the detections made for C3 targets with expectations we achieved a $80\%$ success rate. We are therefore confident that we understand the noise characteristics in K2, and with the updated prescription for the shot noise we are in a good position to propose targets for SC observations in future campaigns.
Campaigns 13 and 18 are especially interesting because we may here complement the seismic analysis with interferometry, and several members of the Hyades are predicted to show detectable oscillations.
We project that by the end of C18 (if selected) we shall have of the order $388$ targets for which a seismic analysis can be accomplished. 
From these we may calibrate seismic scaling relations, especially using the targets with independent constraints from interferometry or precise parallaxes. This is essential to seismic galactic archaeology studies, which rely on such scaling relations. While Gaia will become important for such calibrations there is a strong reciprocity, because the results from our asteroseismic analysis can be used to calibrate the Gaia stellar classification pipeline --- something that has already been requested by the Gaia team. 
With the sampling along the ecliptic these will allow us to study chemical evolution of the solar neighbourhood and place constraints on the age-metallicity relation of nearby field stars. The inferences drawn from K2 will further complement those from TESS, whose observing fields in the baseline mission largely miss the ecliptic.


\acknowledgments
\footnotesize

Funding for this Discovery mission is provided by NASA's Science Mission Directorate.
The authors acknowledge the dedicated team behind the \kp and K2 missions, without whom this work would not have been possible.
M.N.L. acknowledges the support of The Danish Council for Independent Research | Natural Science (Grant DFF-4181-00415).
M.N.L. was partly supported by the European Community's Seventh Framework Programme (FP7/2007-2013) under grant agreement no. 312844 (SpaceInn), which is gratefully acknowledged.
Funding for the Stellar Astrophysics Centre (SAC) is provided by The Danish National Research Foundation (Grant DNRF106). The research was supported by the ASTERISK project (ASTERoseismic Investigations with SONG and \kp) funded by the European Research Council (Grant agreement no.: 267864).
W.J.C. and G.R.D acknowledge the support of the UK Science and Technology Facilities Council (STFC). 
V.S.A. and T.R.W. acknowledges support from VILLUM FONDEN (research grant 10118).
S.B. acknowledges partial support of NASA grant NNX13AE70G and NSF grant AST-1514676.
D.W.L. acknowledges partial support from the \kp mission via Cooperative Agreement NNX13AB58A with the Smithsonian Astrophysical Observatory.
D.H. acknowledges support by the Australian Research Council's Discovery Projects funding scheme (project number DE140101364) and support by the NASA grant NNX14AB92G issued through the Kepler Participating Scientist Program.
This research made use of Astropy, a community-developed core Python package for Astronomy (Astropy Collaboration, 2013).
This research has made use of the SIMBAD database and VizieR access tool, operated at CDS, Strasbourg, France.
This publication makes use of data products from the Two Micron All Sky Survey, which is a joint project of the University of Massachusetts and the Infrared Processing and Analysis Center/California Institute of Technology, funded by the National Aeronautics and Space Administration and the National Science Foundation.


\small
\bibliography{MasterBIB}
\label{lastpage}
\end{document}